\documentclass[apjl]{emulateapj}
\usepackage{epsfig}
\usepackage{verbatim}
\usepackage{natbib}
\usepackage{apjfonts}

\newcommand{\ha}{{H$\alpha$+[N{\sc ii}]~}}

\begin{document}

\shortauthors{TREMBLAY ET AL.}
\shorttitle{{\it HST} EMISSION LINE IMAGING OF RADIO GALAXIES}

\title{{\it HST}~/~ACS Emission Line Imaging of Low Redshift 3CR
Radio Galaxies I: The Data\altaffilmark{1}}

\slugcomment{Accepted for publication in ApJS}

\author{Grant R.~Tremblay\altaffilmark{2,4}}
\author{Marco Chiaberge\altaffilmark{2,3}}
\author{William B.~Sparks\altaffilmark{2}}
\author{Stefi A.~Baum\altaffilmark{4}}
\author{Mark G.~Allen\altaffilmark{5}}
\author{David J.~Axon\altaffilmark{4}}
\author{Alessandro Capetti\altaffilmark{6}}
\author{David J.~E.~Floyd\altaffilmark{7}}
\author{F.~Duccio Macchetto\altaffilmark{2}}
\author{George K.~Miley\altaffilmark{7}}
\author{Jacob Noel-Storr\altaffilmark{4}}
\author{~~~~~~Christopher P.~O'Dea\altaffilmark{4}}
\author{Eric S.~Perlman\altaffilmark{9}}
\author{Alice C.~Quillen\altaffilmark{10}}

\altaffiltext{1}{Based  on  observations  made  with  the NASA/ESA  {\it  Hubble  Space
Telescope}, obtained  at the Space Telescope  Science Institute, which
is  operated  by  the  Association  of Universities  for  Research  in
Astronomy, Inc., under NASA  contract 5-26555}

\altaffiltext{2}{Space  Telescope Science  Institute, 3700  San Martin
Drive, Baltimore, MD 21218; grant@stsci.edu}

\altaffiltext{3}{INAF---Istituto di Radioastronomia, Via
P.~Gobetti 101, Bologna I-40129, Italy}

\altaffiltext{4}{Rochester Institute
of Technology, 84 Lomb Memorial Drive, Rochester, NY 14623; grant@astro.rit.edu}

\altaffiltext{5}{Centre de Donn\'{e}es Astronomique, 11 Rue
de l'Universite, 67000 Strasbourg, France}

\altaffiltext{6}{INAF---Osservatorio  Astronomico  di  Torino,  Strada
Osservatorio 20, 10025 Pino Torinese, Italy}

\altaffiltext{7}{Las Campanas Observatory, Observatories of the Carnegie Institute of Washington,
Casilla 601, La Serena, Chile}

\altaffiltext{8}{Leiden Observatory, P.O. Box 9513, NL-2300 RA
Leiden, The Netherlands}

\altaffiltext{9}{Physics and Space Sciences Department, Florida
Institute of Technology, 150 West University Boulevard,
Melbourne, FL 32901}

\altaffiltext{10}{Department  of Physics  and Astronomy,  University of
Rochester, 600 Wilson Boulevard, Rochester, NY 14627}

\begin{abstract}
We present 19  nearby ($z<0.3$) 3CR radio galaxies  imaged at low- and
high-excitation as  part of  a Cycle 15  {\it Hubble  Space Telescope}
snapshot survey  with the Advanced  Camera for Surveys.   These images
consist  of exposures  of the  H$\alpha$ (6563  \AA, plus  [N{\sc ii}]
contamination)  and  [O{\sc  iii}]$\lambda$5007 emission  lines  using
narrow-band linear ramp filters  adjusted according to the redshift of
the target.  To facilitate continuum subtraction, a single-pointing 60
s line-free  exposure was taken with a  medium-band filter appropriate
for the target's redshift.  We discuss the steps taken to reduce these
images independently of the  automated recalibration pipeline so as to
use more recent ACS flat-field data as well as to better reject cosmic
rays.   We describe the  method used  to produce  continuum-free (pure
line-emission) images, and present these images along with qualitative
descriptions  of the  narrow-line region  morphologies we  observe. We
present  \ha and [O{\sc  iii}] line  fluxes from  aperture photometry,
finding the values to fall expectedly on the redshift-luminosity trend
from a past {\it HST}/WFPC2 emission line study of a larger, generally
higher  redshift subset  of the  3CR.   We also  find expected  trends
between emission line  luminosity and total radio power,  as well as a
positive correlation between the size  of the emission line region and
redshift.  We discuss the  associated interpretation of these results,
and conclude with a summary of future work enabled by this dataset.
\end{abstract}


\section{Introduction}
Characteristically intense  nuclear emission in  radio galaxies, along
with  the  highly collimated  jets  powered  by  accretion onto  their
central  engines,   can  influence  star   formation,  excitation  and
ionization of the ISM, as well  as provide kinetic stresses on the hot
X-ray  emitting  coronal  gas   that  pervades  clusters  of  galaxies
\citep{quillen99,  mcnamara00,  blanton01, reynolds02,  ruszkowski02}.
Narrow-band imaging surveys  of radio galaxies at redshifts  $z > 0.6$
were among  the first studies  to establish such a  connection, having
shown the optical line-emitting gas  ($T \sim 10^4$ K) near the nuclei
of  these  objects to  spatially  align with  the  radio  jet axes  on
kiloparsec scales \citep{fosbury86,hansen87,baum88,mccarthy88,baum90}.
This  so-called  `alignment  effect'  has been  attributed  to  shocks
induced by  propagation of the  radio jet, as well  as photoionization
from  the  AGN  (e.g., \citealt{baum89a,  mccarthy93,best00}).   While
these shocks can  trigger star formation along the  regions of the ISM
excited into  line emission  \citep{chambers87}, more recent  work has
suggested  that  feedback  from  the  AGN  may also  play  a  role  in
ultimately quenching the star formation  rate (SFR) in the late stages
of  the host galaxy's  evolution, thereby  ushering its  rapid passage
from    the   `blue    cloud'   to    the   `red    sequence'   (e.g.,
\citealt{cowie96,bell04,faber07},  and references  therein).   Work is
ongoing in  reconciling the seemingly competing roles  of AGN feedback
in  both   inducing,  then  possibly  truncating   star  formation  at
successive      stages      of      galactic     evolution      (e.g.,
\citealt{silk98,fabian99,dimatteo05,hopkins05,silverman08}).
Regardless,  it  is clear  that  the AGN  and  its  host galaxy  share
codependent    evolutionary    paths   (e.g.,    \citealt{magorrian98,
  ferrarese00,gebhardt00}), and that studies of  the ionized nuclear gas in radio
galaxies,  as unique  probes of  the early  universe, will  be  key to
understanding  this relationship  on size  and time  scales  small and
large.

The {\it Hubble Space Telescope} ({\it HST}) has undertaken systematic
surveys   of  the   3CR  catalog   of   radio  galaxies   in  the   UV
\citep{allen02},   optical   \citep{martel99,dekoff00,privon08},   and
near-infrared  \citep{madrid06,donzelli07,tremblay07,floyd08}.   These
programs, in concert with  a robust array of ground-based observations
of  the 3CR  in  nearly  all wavelength  regimes,  have established  a
uniform  database of  cross-spectrum  imaging and  spectroscopy for  a
sample that is nearly complete  with redshift and flux limit, unbiased
with  respect  to  optical/IR  wavelengths,  and  containing  galaxies
exhibiting a  variety of intrinsic  characteristics.  The completeness
and  diversity  of the  sample  has  enabled  studies into  radio-loud
unification  models and  the dichotomy  between relatively  low power,
edge-darkened Fanaroff and Riley  class I radio galaxies (hereafter FR
I, \citealt{fanaroff74}) and higher power, edge-brightened FR~II radio
galaxies \citep{chiaberge00a, chiaberge02}.  {\it HST} observations of
the  3CR have  also led  to  discoveries of  new optical  and IR  jets
\citep{floyd06}, a  trend between  nuclear dusty disk  inclination and
host galaxy isphotal shapes  \citep{tremblay07} and face on disks with
optical jets \citep{sparks00}.   Jet/disk orientations were studied at
unprecedented            spatial           resolution           (e.g.,
\citealt{schmitt02,verdoeskleijn05,tremblay06},    and    references   therein).
\citet{allen02} also  found complex UV morphologies  unlike those seen
in quiescent hosts, indicative of ongoing star formation.

In this paper we present  narrow-band emission line observations of 19
nearby  ($z  < 0.3$)  3CR  radio galaxies  imaged  at  low- and  high-
excitation  with the  Advanced Camera  for Surveys  (ACS)  aboard {\it
  HST}.  These images  of the warm optical line  (H$\alpha$ and [O{\sc
    iii}]$\lambda  5007$) emitting gas  present in  the nuclei  of the
radio galaxies in our sample offer the highest sensitivity and spatial
resolution yet available in the  {\it HST} 3CR database.  This enables
detailed studies  of shocked  and star-forming regions  of the  ISM in
relation to radio jets, providing  the already rich 3CR dataset with a
larger framework  from which to study the  various phenomena discussed
above.

We organize this paper as follows.  In \S2 we describe these {\it HST}
observations  in detail,  and in  \S3 we  discuss the  associated data
reduction.   In  \S4 we  present  the  \ha (low-excitation)  and
[O{\sc  iii}]  (high-excitation)  emission  line images,  as  well  as
qualitative  descriptions  of each.   We  also  provide emission  line
luminosity results  and an  associated discussion.  We  summarize this
work  and  discuss future  studies  enabled  by  this new  dataset  in
\S5. Throughout this  paper we use $H_0 =  71$ km s$^{-1}$ Mpc$^{-1}$,
$\Omega_M = 0.27$, and $\Omega_{\Lambda} = 0.73$.

\section{Sample Selection \& Observations}

In this paper we present 19 radio galaxies from  the
low-redshift  ($z < 0.3$)  extragalactic subset  of the  Revised Third
Cambridge  Catalog  (3CR,  \citealt{bennett62a,bennett62b,spinrad85}).
3CR sources  are selected by radio  flux density at 178  MHz, at which
extended, unbeamed lobes are  detected irrespective of the radio jet's
orientation  with  respect to  the  line  of  sight.  The  sample  is
therefore free from orientation bias and
is  nearly complete with  redshift, providing  a strong  framework for
statistical  analysis  important  for  studies of  (e.g.)   radio-loud
unification models.   Moreover, the sample consists  of radio galaxies
that  exhibit a  wide  variety of  characteristics  (size, shape,  jet
morphology, radio  luminosity, FR type, dust  content, etc.), allowing
for  physical comparisons  to be  made among  properties  intrinsic to
various species  of radio-loud AGN. We select  the low-redshift subset
of the 3CR  so as to achieve the  highest possible spatial resolution,
in addition to requiring the  H$\alpha$ line to be within the redshift
range of high sensitivity to provide uniform line ratio mapping. 


Of the 116  extragalactic 3CR sources with $z <  0.3$, 98 were awarded
for  observation in  the  Cycle  15 ACS  snapshot  program 10882  (PI:
Sparks).   The remaining  18 targets  that were  not requested  in our
proposal  were excluded for  various reasons,  mainly due  to emission
line  data already  being  present  in the  archive  for that  object.
Ultimately, 20 of  our awarded targets were observed  prior to failure
of ACS side  two electronics in January 2007.   Of these observations,
19 were  successful.  Our  observation of 3C~371  was carried  out but
unsuccessful  due to failure  of guide  star acquisition,  resulting in
unusable data. The sample consists of 4 FR~I radio galaxies, 12 FR~II
radio galaxies, and 3  targets exhibiting compact steep spectrum (CSS)
or undiscernible radio morphology.

\begin{figure}
\plotone{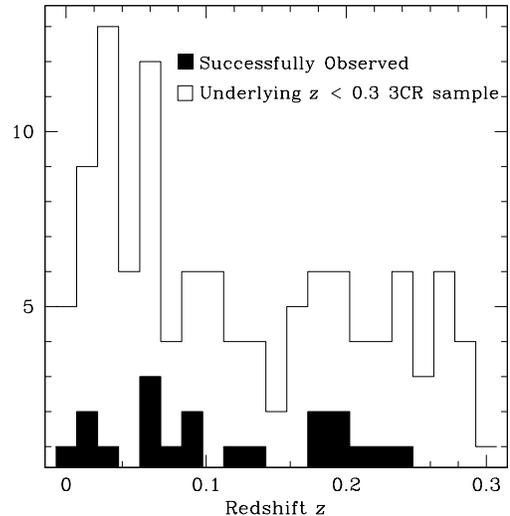}
\caption{Redshift  distribution  of  the  low  redshift  ($z  <  0.3$)
  extragalactic subset of the 3CR sample (116 galaxies), compared with
  the  distribution  of  targets  successfully imaged  (prior  to  ACS
  failure) as part of the ACS snapshot program presented in this paper
  (solid,  19  galaxies).   While  this  dataset  represents  a  small
  fraction of the underlying sample,  it spans nearly the entire range
  of  redshift.  Collectively, the  vast  majority  of the  underlying
  sample has been imaged as part of past {\it HST} programs in the UV,
  optical,  and near  IR. See  \S1 for  a summary  of these  {\it HST}
  programs.}
\label{fig:redshiftfig}
\end{figure}

While  only  a small  subset  of  our  initially proposed  sample  was
observed,  the  nature  of   {\it  HST}'s  snapshot  mode  means  that
observations were scheduled by convenience with respect to {\it HST}'s
other  programs at the  time, and  not by  any selection  effect.  Our
group  of  galaxies  is  therefore  as unbiased  as  the  3CR  itself.
Moreover, the 19 galaxies  successfully observed span nearly the whole
redshift range ($0.0075 <  z < 0.224$, see Fig.~\ref{fig:redshiftfig})
of the underlying sample and, as  we will discuss in more detail in \S
4,  includes  galaxies  exhibiting  a  wide  array  of  emission  line
characteristics.

Our 19 successful observations consist of ACS Wide Field Channel (WFC)
exposures  of  the \ha  and  [O{\sc iii}]$\lambda$5007  emission
lines using  narrow-band (2\% bandpass) linear ramp  filters (LRFs) at
wavelengths  adjusted according  to the  redshift of  the  target (see
Table~\ref{tab:tab1}).  The LRFs  aboard ACS feature effective central
bandwidths that vary  as the filter is rotated  across the FOV.  While
the  LRFs are  useful in  providing the  observer  narrow-band imaging
capabilities  at a  wide variety  of central  wavelengths, there  is a
drawback in  that the target  is imaged monochromatically only  over a
small region  ($\sim 40\arcsec \times 80\arcsec$)  of the instrument's
FOV.  In planning  the observations for this dataset,  we ensured that
the target  galaxy nuclei always  corresponded to the very  center of
this region.

\begin{figure*}
\plotone{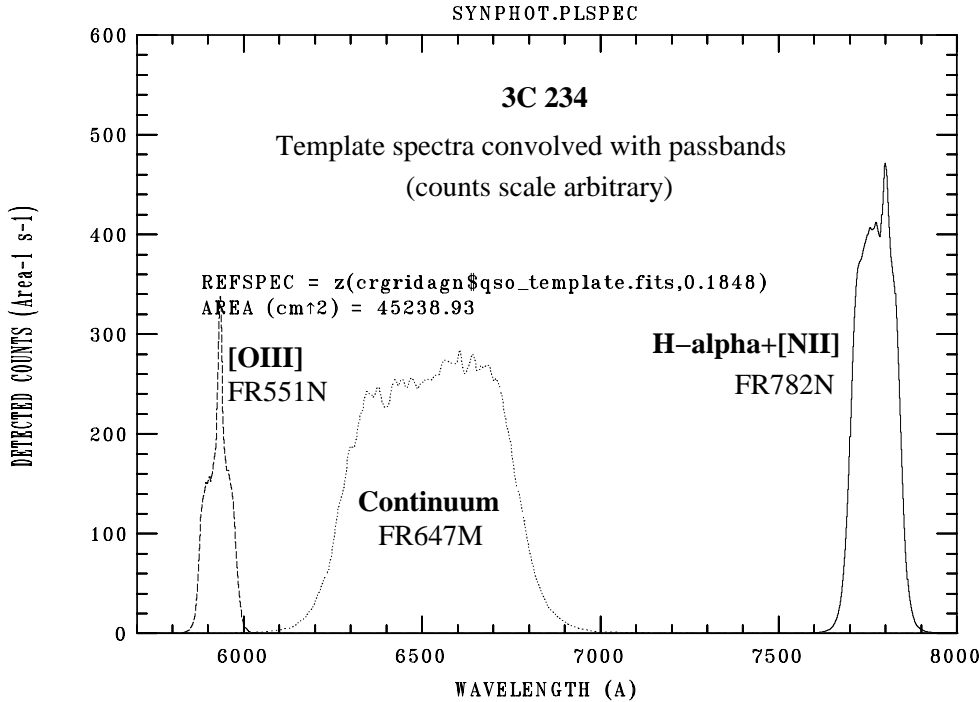}
\caption{A  representative example of  our filter  selection strategy.
  We  have convolved  {\it template}  LINER (chosen  only  for example
  H$\alpha$ and [N{\sc ii}] lines)  and QSO (for [O{\sc iii}]) spectra
  through  the filters  chosen for  our observations  of 3C~234  ($z =
  0.18$) as an example. The  spectra are used only as an illustration,
  and are not meant for  quantitative discussion.  The counts scale is
  therefore arbitrary.   See \S2 for  details on our  filter selection
  strategy. }
\end{figure*}

With each target  fully observed in one orbit, an  exposure time of $2
\times 200$  s and  $2 \times 250$  s was  used for the  H$\alpha$ and
       [O{\sc  iii}]  lines, respectively,  with  a  two point  dither
       pattern (of line spacing $0\arcsec.145$) to aide cosmic ray
       and  hot  pixel  rejection.   So  as  to  facilitate  continuum
       subtraction  from the  emission  line images,  a  60 s,  single
       pointing exposure  was taken with a medium  (9\%) bandpass ramp
       filter centered  at rest-frame 5500  \AA, covering a  region of
       the optical  spectrum absent of  significant contamination from
       line emission.  Subarrays were  used to increase the efficiency
       of these  observations, allowing 4  images to be stored  in the
       ACS  buffer at  once.  The  WFC on  ACS consists  of  two $2048
       \times  4096$ pixel  CCDs (WFC1  and WFC2),  each with  a plate
       scale of $0\arcsec.049$ per pixel  and a field of view (FOV) of
       $202\arcsec \times 202\arcsec$ (further details may be found in
       the ACS Instrument Handbook\footnote{Boffi, F.R., et al.  2007,
         ``ACS   Instrument   Handbook'',   Version   8.0,   (Baltimore:
         STScI).}).  See  Table \ref{tab:tab1} for  an observation log
       related to this program.

\begin{figure*}
\plotone{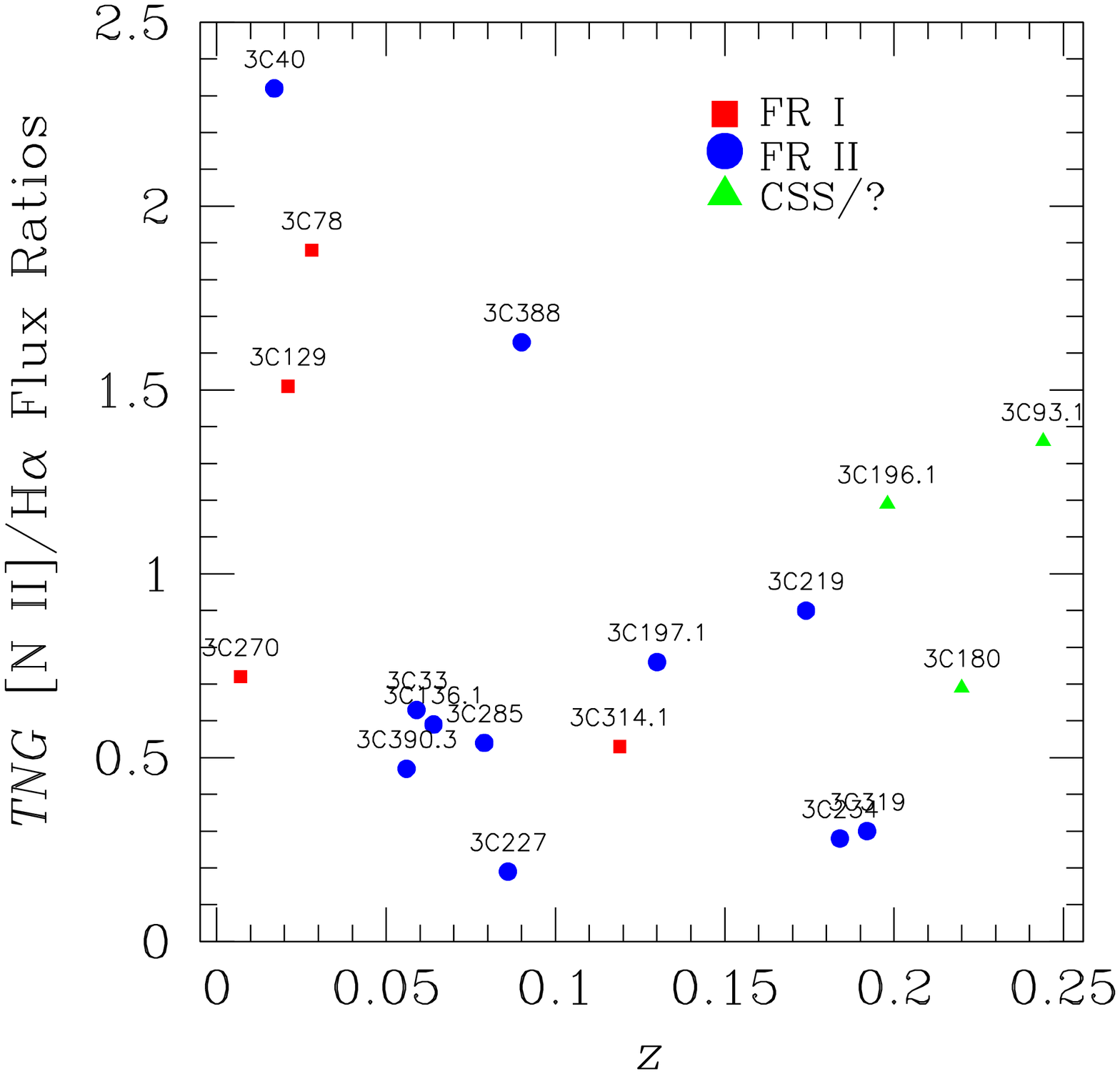}
\caption{An illustrative  example of the  degree to which  [N{\sc ii}]
  flux  varies relative  to  H$\alpha$ among  the  3CR radio  galaxies
  discussed in  this paper.  Here  we plot de-reddened flux  ratios of
  the   diagnostic   line   [N{\sc  ii}]$\lambda6584$   to   H$\alpha$
  vs.~redshift.   The flux  ratios are  from the  ground-based optical
  spectroscopy performed by \citet{buttiglione09} using the Telescopio
  Nazionale Galileo (TNG).   The results from that work  show that the
  relative contributions of [N{\sc ii}] and H$\alpha$ vary widely from
  source-to-source, and  that [N{\sc ii}] appears to  dominate in most
  high-excitation  galaxies  (HEG,   see  Table  \ref{tab:tab2}).  Red
  squares represent the  four FR~Is in our sample,  while blue circles
  are FR~IIs and green  triangles represent those objects exhibiting a
  CSS or  unclassified radio morphology. 3C~132 is  excluded from this
  plot as no TNG spectroscopy was available for this object.}
\label{fig:ratios}
\end{figure*}

It is important to note  that, for all ``H$\alpha$ imaging'' presented
in  this paper,  the bandpass  used  was contaminated  by [N{\sc  ii}]
emission at 6548 and 6583 \AA.  References to H$\alpha$ will therefore
read ``\ha'' throughout  this paper. Though it is  beyond the scope of
this work  to subtract this  contamination, [N{\sc ii}] flux  does not
impact the results  presented in this paper in any  way as we present
all corellations  with respect to  \ha and
never  H$\alpha$ alone.   In  follow-up papers  providing analysis  of
these data, however, new  ground-based optical spectroscopy of the 3CR
by  \citet{buttiglione09}   may  be   used  to  better   quantify  the
contribution  of [N{\sc  ii}]  through each  H$\alpha$ bandpass.   The
Buttiglione spectra,  obtained using the  Telescopio Nazionale Galileo
(TNG), may  be convolved with the  spectral response of  the {\it HST}
ACS ramp filters,  providing a constraint on the  relative strength of
H$\alpha$ and  [N{\sc ii}] in  the filter throughput.  In  doing this,
one must take into account the fact that the TNG spectra are extracted
from a 2\arcsec  $\times$ 2\arcsec region, while many  of our galaxies
(e.g., 3C~33) are far more extended.  In Fig.~\ref{fig:ratios} we plot
the  de-reddened   flux  ratios  from   \citet{buttiglione09}  of  the
diagnostic line  [N{\sc ii}]$\lambda6584$ to  H$\alpha$, vs.~redshift,
for  each galaxy in  our sample  (with the  exception of  3C~132). The
results  from \citet{buttiglione09},  some  of which  are apparent  in
Fig.~\ref{fig:ratios}, show that  the relative contributions of [N{\sc
    ii}]  and H$\alpha$  vary widely  from source-to-source,  and that
[N{\sc ii}] appears to dominate in most high-excitation galaxies (HEG,
see Table \ref{tab:tab2}).\\

\begin{figure*}
\plotone{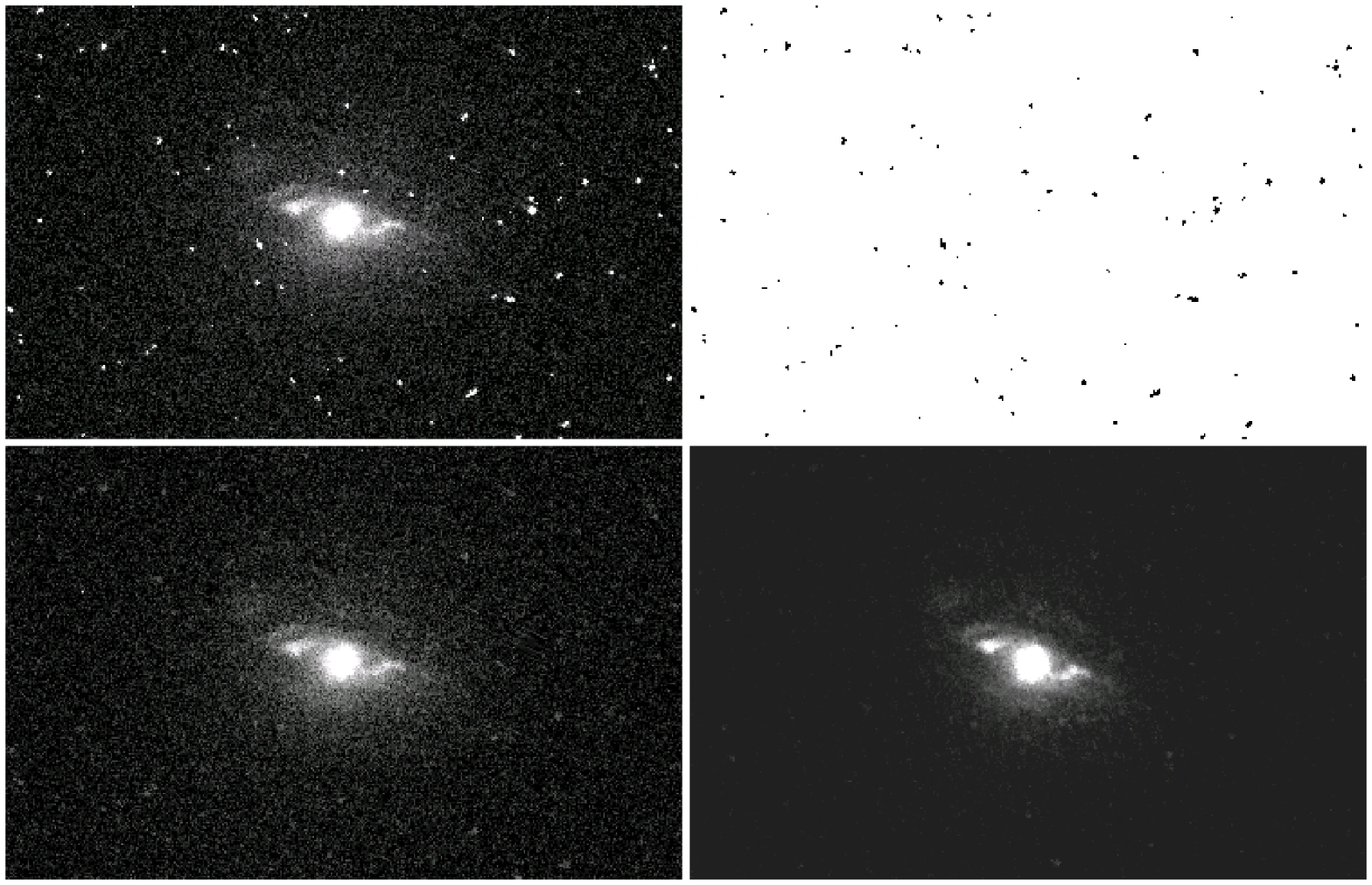}
\caption{A representative example of our iterative masking process for
  cleaning each image of cosmic rays and hot pixels.  ({\it Top left})
  Calibrated (by \texttt{calacs}, see \S3.1) single-pointing \ha
  exposure of 3C~33, one of two exposures that are combined during the
  \texttt{multidrizzle}  process   to  produce  a   single  image  (as
  described in  \S3.3).  ({\it Top right})  One of the two  CR and hot
  pixel masks created  for the exposure presented at  ({\it Top left})
  by the \texttt{cosmicrays} routine, as discussed in \S3.2. This mask
  was used to specify locations on the exposure in ({\it Top left}) at
  which  to   linearly  interpolate   across  bad  pixel   regions  by
  neighboring   good  pixels,   as  described   in  \S3.2.    In  this
  representation, the  bad pixels are displayed as  black. ({\it Bottom
    left})  The  same  exposure  as  in ({\it  Top  left})  after  two
  iterations of the  CR/hot pixel cleaning process with  the bad pixel
  masks.  ({\it  Bottom right}) The combined  drizzled \ha image
  of  3C~33,  using  the  two  dithered exposures  (one  of  which  is
  presented at {\it Bottom left})  that have been cleaned prior to the
  use of \texttt{multidrizzle} by our iterative masking process. These
  images  are significantly cleaner  than those  provided by  the OTFR
  pipeline, particularly in the  case of the single-pointing continuum
  images. See \S3 for more information.  }
\label{fig:CRfig}
\end{figure*}

\section{Data Reduction}

Here  we  describe the  steps  taken to  produce  data  that has  been
calibrated, cleaned of  cosmic rays (CRs) and hot  pixels, and finally
subtracted of  contamination from  continuum emission.  The  {\it HST}
On-the-Fly Recalibration (OTFR)  pipeline utilizes pre-launch (ground)
flats  in the  reduction  of ACS  LRF  images, the  use  of which  can
contribute  upwards  of 7\%  uncertainty  to  reduced  data given  the
evolution of  the instrument in  the years since launch.   By reducing
each  image  ``manually'',  we  are  able to  avoid  introducing  this
uncertainty,  as  well  as  implement  a more  proactive  approach  to
cleaning the images of CRs and hot pixels, particularly in the case of
the single-pointing continuum images  for which automated CR rejection
via combination of dithered images is not possible.  For these reasons
the images presented in this  paper have been reduced independently of
the OTFR pipeline, in a process we describe below.


\begin{deluxetable*}{lccccclcr}
\tabletypesize{\scriptsize}
\tablecaption{Observation Log}
  \tablewidth{0pc}
  \tablehead{
    \colhead{Source} &
    \colhead{$z$} &
    \colhead{Obs. Date (UT)} &
    \colhead{$\alpha$ (J2000.0)} &
    \colhead{$\delta$ (J2000.0)} &
    \colhead{Filters} &
    \colhead{Line/Cont.}  &
    \colhead{$\lambda$ (\AA)} &
    \colhead{Exp. Time (s)} \\
    \colhead{(1)} & \colhead{(2)} & \colhead{(3)} &
    \colhead{(4)} & \colhead{(5)} & \colhead{(6)} &
    \colhead{(7)} & \colhead{(8)} & \colhead{(9)}}
  \startdata  
3C~33.0  & 0.0597 & 20 Aug 2006 & 01 08 52.8 & +13 20 14 &  FR647M &  Cont. & 5824 & $1 \times 60$  \\
&  & & & &  FR716N &  H$\alpha$ & 6950 & $2 \times 200$   \\
&  & & & &  FR551N &  [O{\sc iii}] & 5302 & $2 \times 250$   \\
3C~40.0  & 0.0180 & 17 Sep 2006 & 01 20 34.0 & -01 20 34 & FR647M & Cont. & 5593 & $1 \times 60$  \\
 &  &  & &  & FR656N &  H$\alpha$ & 6674 & $2 \times 200$   \\
 &  &  &  &  &  FR505N &  [O{\sc iii}] & 5092& $2 \times 250$   \\
3C~78.0  & 0.0286 & 13 Sep 2006 & 03 08 26.2 & +04 06 39 &  FR647M &  Cont. & 5654 & $1 \times 60$   \\
 &  & &  &  &  FR656N &  H$\alpha$ & 6747 & $2 \times 200$   \\
 &  &  &  &  &  FR505N &  [O{\sc iii}] & 5147 & $2 \times 250$   \\
3C~93.1  & 0.2430 & 8 Sep 2006  & 03 48 46.9 & +33 53 15 &  FR647M & Cont. & 6842 & $1 \times 60$  \\
 &  &  &  &  &  FR853N &  H$\alpha$ & 8146 & $2 \times 200$   \\
 &  &  &  &  &  FR601N &  [O{\sc iii}] & 6228 & $2 \times 250$   \\
3C~129.0 & 0.0208 & 25 Oct 2006 & 04 49 09.1 & +45 00 39 & FR647M & Cont. & 5615 & $1 \times 60$   \\
 &  &  &  &  &  FR656N &  H$\alpha$ & 6701 & $2 \times 200$   \\
 &  &  &  &  &  FR505N &  [O{\sc iii}] & 5112 & $2 \times 250$   \\
3C~132.0 & 0.2140 & 20 Aug 2006 & 04 56 43.0 & +22 49 22 &  FR647M &  Cont. & 6677 & $1 \times 60$  \\
 &  &  & &  &  FR782N &  H$\alpha$ & 7967 & $2 \times 200$   \\
 &  &  & &  &  FR601N &  [O{\sc iii}] & 6078 & $2 \times 250$   \\
3C~136.1 & 0.0640 & 4 Nov 2006  & 05 16 03.1 & +24 58 25 &  FR647M &  Cont. & 5852 & $1 \times 60$  \\
& &  &  &  &  FR716N &  H$\alpha$ & 6983 & $2 \times 200$   \\
& &  &  &  &  FR551N &  [O{\sc iii}] & 5327 & $2 \times 250$   \\
3C~180.0 & 0.2200 & 11 Nov 2006 & 07 27 04.5 & -02 04 42 &  FR647M & Cont. & 6710 & $1 \times 60$  \\
 &  &  &  &  &  FR782N &  H$\alpha$ & 8006 & $2 \times 200$   \\
 &  &  &  &  &  FR601N &  [O{\sc iii}] & 6108 & $2 \times 250$   \\
3C~196.1 & 0.1980 & 3 Dec 2006  & 08 15 27.8 & -03 08 27 &  FR647M &  Cont. & 6588 & $1 \times 60$ \\
 &  &  &  &  &  FR782N &  H$\alpha$ & 7862 & $2 \times 200$   \\
 &  &  &  &  &  FR601N &  [O{\sc iii}] & 5998 & $2 \times 250$   \\
3C~197.1 & 0.1280 & 16 Sep 2006 & 08 21 33.6 & +47 02 37 &  FR647M &  Cont. & 6215 & $1 \times 60$  \\
 &  &  &  &  &  FR716N &  H$\alpha$ & 7416 & $2 \times 200$   \\
 &  &  &  &  &  FR551N &  [O{\sc iii}] & 5657 & $2 \times 250$   \\
3C~219.0 & 0.1744 & 5 Dec 2006  & 09 21 08.6 & +45 38 57 &  FR647M  &   Cont. & 6457 & $1 \times 60$  \\
 &  &  &  &  &  FR782N &  H$\alpha$ & 7704 & $2 \times 200$   \\
 &  &  &  &  &  FR601N &  [O{\sc iii}] & 5878 & $2 \times 250$   \\
3C~227.0 & 0.0858 & 22 Jan 2007 & 09 47 45.1 & +07 25 21 &  FR647M &  Cont. & 5973 & $1 \times 60$  \\
& & & & &  FR716N &  H$\alpha$ & 7127 & $2 \times 200$   \\
& & & & &  FR551N &  [O{\sc iii}] & 5437 & $2 \times 250$   \\
3C~234.0 & 0.1848 & 15 Dec 2006 & 10 01 49.5 & +28 47 09 &  FR647M &   Cont. & 6512 & $1 \times 60$ \\
 &  &  &  &  &  FR782N &  H$\alpha$ & 7770 & $2 \times 200$   \\
 &  &  &  &  &  FR601N &  [O{\sc iii}] & 5928 & $2 \times 250$   \\
3C~270.0 & 0.0075 & 25 Dec 2006 & 12 19 23.2 & +05 49 31 &  FR647M & Cont. & 5541 & $1 \times 60$ \\
         &       &               & &  &  FR656N &  H$\alpha$ & 6612 & $2 \times 200$   \\
 &  &  &  &  &  FR505N &  [O{\sc iii}] & 5044 & $2 \times 250$   \\
3C~285.0 & 0.0794 & 11 Jan 2007 & 13 21 17.8 & +42 35 15 &  FR647M &  Cont. & 5934 & $1 \times 60$  \\
&  &  & & &  FR716N &  H$\alpha$ & 7081 & $2 \times 200$   \\
&  &  & & &  FR551N &  [O{\sc iii}] & 5402 & $2 \times 250$   \\
3C~314.1 & 0.1197 & 30 Dec 2006 & 15 10 22.5 & +70 45 52 &  FR647M &   Cont. & 6154 & $1 \times 60$  \\
 &  &  &  &  &  FR716N &  H$\alpha$ & 7343 & $2 \times 200$   \\
 &  &  &  &  &  FR551N &  [O{\sc iii}] & 5602 & $2 \times 250$   \\
3C~319.0 & 0.1920 & 12 Nov 2006 & 15 24 05.5 & +54 28 15 &  FR647M &  Cont. & 6556 & $1 \times 60$ \\
 &  &  &  &  &  FR782N &  H$\alpha$ & 7823 & $2 \times 200$   \\
 &  &  &  &  &  FR601N &  [O{\sc iii}] & 5968 & $2 \times 250$   \\
3C~388.0 & 0.0917 & 29 Dec 2006 & 18 44 02.4 & +45 33 30 &  FR647M &  Cont. & 6004 & $1 \times 60$   \\
& & & & &  FR716N &  H$\alpha$ & 7164 & $2 \times 200$   \\
& & & & &  FR551N &  [O{\sc iii}] & 5466 & $2 \times 250$   \\
3C~390.3 & 0.0561 & 30 Dec 2006 & 18 42 09.0 & +79 46 17 &  FR647M &  Cont. & 5808 & $1 \times 60$ \\
 &  &  &  &  &  FR716N &  H$\alpha$ & 6930 & $2 \times 200$   \\
 &  &  &  &  &  FR551N &  [O{\sc iii}] & 5287 & $2 \times 250$   \cr
  \enddata
  \tablecomments{
The 19 3CR radio galaxies observed as part of the {\it HST} Cycle 15 SNAP program 10882 (PI: Sparks),
listed by source name in ascending order.
    (1) 3CR source name;
    (2) redshift;
    (3) observation date;
    (4) right ascension (J2000.0, in hours, minutes, and seconds);
    (5) declination (J2000.0, in degrees, arcminutes, and arcseconds);
    (6) ACS ramp filters used, listed for the 60 s continuum, the 400 s H$\alpha$,
        and the 500 s [O{\sc iii}]$\lambda$5007 observations, respectively;
    (7) Emission line (or continuum) observed, corresponding to filter configuration;
    (8) Redshift-adjusted Wavelength of emission line being observed, corresponding to the
wavelength to which the ramp filter has been configured, with 2\%
bandpass for the H$\alpha$ and [O{\sc iii}] exposures, and 9\% bandpass for the
continuum exposures;
    (9) Exposure time in seconds. The continuum images are single
exposures, while the emission line observations consist of two exposures in a two-point
dither pattern. See \S2 for more details on the sample selection and observations for this program.
Also part of this program was 3C~371, the observation of which failed due to failure 
of guidestar aquisition. It is therefore excluded from
presentation and analysis in this paper.
See Table \ref{tab:tab2} for a summary of the optical properties of the above sample, and
Table \ref{tab:tab3} for a summary of its radio properties.
}
\label{tab:tab1}
\end{deluxetable*}

\subsection{Calibration}

The calibration stage  begins with the raw exposures  and corrects for
flat-fielding, bias and dark  levels, as well as absolute sensitivity.
Each of  these corrections is  performed by the  IRAF\footnote{IRAF is
  distributed by the National  Optical Astronomy Observatory, which is
  operated  by  the  Association   of  Universities  for  Research  in
  Astronomy  (AURA)  under  cooperative  agreement with  the  National
  Science      Foundation.}       routine     \texttt{calacs}      (in
\texttt{stsdas.hst\_calib.acs}).  So as  to avoid the uncertainty that
would otherwise have been introduced by the use of pre-launch flats in
the OTFR  pipeline, we updated  the {\sc pfltfile} header  keyword for
each raw image  to specify that the newest  available flat be utilized
for each \texttt{calacs} run.

In  selecting the  appropriate new  flat to  use (in  lieu of  the old
ground flats  used by the OTFR  pipeline) we employ  the same strategy
that was used when the original ground flats for the ramp filters were
created.   That is,  LRF flats  are  interpolated from  flats for  ACS
narrow, medium, and broad-band (i.e.  non LRF) filters that correspond
most  closely with  the redshift-adjusted  wavelength of  the emission
line  (or continuum) being  imaged.  The  minor difference  in central
wavelength,  throughput,  and bandwidth  of  these  (non LRF)  filters
corresponds  to  negligible differences  in  their corresponding  flat
fields (less than  the $\sim 7$\% uncertainty introduced  in using the
old ground LRF flats).

With  the replacement  flats chosen  on a  case-by-case basis  and the
image headers  updated to  reflect these changes,  the \texttt{calacs}
routine was run  for each image, producing calibrated  FITS files (one
for the continuum exposure, and  two each for the \ha and [O{\sc
    iii}]  exposures) that  have been  flat-fielded and  corrected for
bias,  dark  current,  and  absolute  sensitivity.  It  is  upon  these
calibrated  images  that  cosmic  ray  and  hot  pixel  rejection  was
performed, as we discuss in the following section.

\begin{figure*}
\plotone{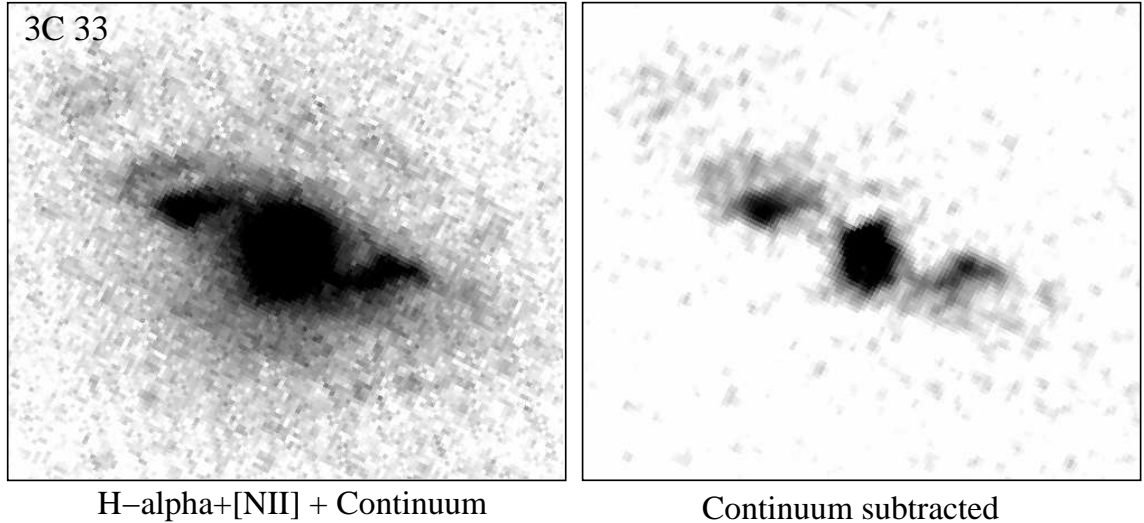}
\caption{A  ``before and  after''  example result  from our  continuum
  subtraction process. ({\it  a}) Reduced FR716N $H\alpha$+[N{\sc ii}]
  image of  3C~33, contaiminated by  galaxy continuum.  ({\it  b}) The
  same  image,  with  galaxy   continuum  subtracted  via  the  method
  discussed in \S3.5}
\label{fig:contsub}
\end{figure*}

\subsection{Cosmic ray and hot pixel rejection}

The  NOAO IRAF  task  \texttt{cosmicrays} (\texttt{noao.imred.crutil})
was run  on copies of the  science frames of  each calibrated exposure
(one for  the continuum image, two  each for the \ha  and [O{\sc iii}]
images) to identify CRs and hot pixels.  The routine identifies cosmic
ray events  by a detection algorithm  based on a  mean flux comparison
within a specified detection window.  A pixel within this window whose
value  is higher than  a user-specified  threshold dependent  upon the
mean value  of its  neighbors is considered  ``bad'', and  flagged for
cleaning by  the routine.  Once the  list of cosmic ray  and hot pixel
candidates  is established over  the whole  image, \texttt{cosmicrays}
replaces flagged pixels with the  mean value of its four neighbors.  A
``cleaned'' output file is then  generated.  We were careful to ensure
that   \texttt{cosmicrays}   did   not  mis-identify   real   features
(particularly galaxy nuclei or other physical, bright features) as CRs
or hot pixels. The routine's  parameters were adjusted if this was the
case.

The  ``cleaned'' output from  \texttt{cosmicrays} is  insufficient for
our needs,  as the  routine will often  simply ``dull'' the  CR rather
than  truly replace  it.   This is  because \texttt{cosmicrays}  flags
pixels  within  a detection  window  that  often contains  neighboring
pixels  that  are also  affected  by the  CR  (and  are therefore  far
brighter than the true background).  In the worst instances, depending
on the  parameters that are chosen,  \texttt{cosmicrays} can sometimes
{\it reconstruct} the very CR it was attempting to mask.

Therefore, we utilized the ``cleaned'' output from \texttt{cosmicrays}
only as an intermediate step towards  the creation of a bad pixel mask
for each  exposure.  To that  end, we divided the  original, uncleaned
copy of the calibrated science  frame with the ``cleaned'' output from
\texttt{cosmicrays}.  The divided image  was then examined by eye, and
blinked with  the uncleaned  frame so as  to qualitatively  assess the
success of  \texttt{cosmicrays} in identifying  offending pixels.  The
bad  pixel mask  was then  created from  the divided  image,  in which
pixels whose  values were above  a specified threshold  dictated those
that were  to be  masked in the  uncleaned image.  This  threshold was
decided qualitatively  for each image  based on our assessment  of the
success of \texttt{cosmicrays} in distinguishing truly bad pixels from
real features.  Often, accepting every flagged pixel  from the divided
image into  the final mask would  result in masking  healthy pixels on
the galaxy itself.   If this appeared to be  the case, the ``allowance
threshold'' was set {\it above}  the values of the incorrectly flagged
pixels in  the divided image, so that  they were not made  part of the
final mask.  We paid particular attention to flagged pixels within the
immediate vicinity of the target galaxy, and ensured on a case-by-case
basis that the pixels being masked were indeed CR events or hot pixels
and not  real features in  the galaxy.  This was  accomplished through
blinking individual  exposures or by comparing the  continuum image to
archival {\it  HST} data  from e.g.~WFPC2 in  the few cases  where the
correct choice was not obvious.

The  mask was  then used  with  the \texttt{fixpix}  task to  linearly
interpolate values for  masked pixels based on those  of their nearest
unmasked neighbors. While  this is in effect similar  to the procedure
employed   by  \texttt{cosmicrays},   it  is   more  robust   in  that
\texttt{fixpix}  is better  able to  replace  a CR  event with  pixels
generally      consistent     with      that      of     the      true
background. \texttt{cosmicrays}, on the  other hand, will often simply
``dull''  the CR  rather than  truly replace  it, as  it  flags pixels
within  a  detection window  that  often  contains neighboring  pixels
affected  by the  CR (and  are therefore  far brighter  than  the true
background).  Moreover,  the masking process allows the  user to check
at  each  step that  no  real feature  on  the  galaxy is  erroneously
altered.   This process  was repeated  several times  (typically three
times)  for  {\it  each}   calibrated  science  frame  (one  continuum
exposure,  two \ha  frames, and  two  [O{\sc iii}]  frames).  At  each
iteration  in  the  process  we  varied the  detection  thresholds  in
\texttt{cosmicrays},  as  well   as  the  parameters  associated  with
creating the subsequent mask, in order  to ensure that all CRs and hot
pixels    were     ultimately    identified    and     cleaned.     In
Fig.~\ref{fig:CRfig} we provide a  representative example of the steps
in our cosmic ray and hot pixel rejection strategy. The upper-left and
lower-right  panels  in the  figure  serve  as  a "before  and  after"
comparison.

In two cases (3C~33 and 3C~78)  a large CR event landed directly on or
near  to the  target galaxy  image  in the  single pointing  continuum
exposure.  In these cases we masked the CR while taking care to affect
the galaxy light distribution as little as possible.  Inevitably, some
error due  to pixel value  interpolation remains on the  final reduced
images  (as the continuum  exposure is  used for  both \ha  and [O{\sc
    iii}] in  the continuum subtraction process,  described in \S3.4).
However, while this error is  difficult to characterize it is unlikely
that it has a significant effect  on the ultimate science value of the
data, both  for morphological studies of the  emission-line region and
for e.g.~flux measurements.

\subsection{Multidrizzle}

The    calibrated,   cleaned    data   were    passed    through   the
\texttt{multidrizzle} routine  with the default  parameters.  The task
calculates  and subtracts a  background sky  value for  each exposure,
searches for  additional bad  pixels not already  flagged in  the data
quality array,  and drizzles the  two exposures into outputs  that are
shifted  and  registered with  respect  to  one  another.  From  these
drizzled exposures a  median image is created, which  is then compared
with original  input images  so as  to reject the  (very few)  CRs not
already cleaned in  our previous CR rejection process,  as we describe
in  the   section  above.   More  information  on   the  specifics  of
\texttt{multidrizzle}  can  be  found in  \citet{koekemoer02}.   Final
output images were left unrotated with respect to north to avoid pixel
interpolation  errors  that might  add  uncertainty  to our  continuum
subtraction and  flux estimation strategies, which we  describe in the
sections below.

\begin{deluxetable*}{lccccccc}
\tabletypesize{\scriptsize}  
\setlength{\tabcolsep}{0.02in}
\tablecaption{Optical Properties}
  \tablewidth{0pc}
  \tablehead{
    \colhead{} &
    \colhead{H$\alpha$ Line Flux} &
    \colhead{[O{\sc iii}] Line Flux} &
    \colhead{log $L_{\mathrm{H}\alpha}$ } &
    \colhead{log $L_{\mathrm{[OIII]}}$ } &
    \colhead{} &
    \colhead{L.A.S$_{\mathrm{H}\alpha}$} \\
    \colhead{Source} &
    \colhead{($\times 10^{-15}$ erg s$^{-1}$ cm$^{-2}$)} &
    \colhead{($\times 10^{-15}$ erg s$^{-1}$ cm$^{-2}$)} &
    \colhead{(erg s$^{-1}$)} &
    \colhead{(erg s$^{-1}$)} &
    \colhead{Ionization Class$^{\mathrm{~a,b}}$} &
    \colhead{(kpc)} \\
    \colhead{(1)} & \colhead{(2)} & \colhead{(3)} &
    \colhead{(4)} & \colhead{(5)} & \colhead{(6)} & \colhead{(7)}}
  \startdata
3C~33.0  & 0.428   & 1.58  & 39.5255 &40.0927  &  HEG &5.142 \\
3C~40.0  & 1.69  & 1.66 &   39.0232 & 39.0154 &  LEG & 1.259  \\
3C~78.0  & 2.18    & 2.15   &39.5719  &39.5659  & \nodata &2.208   \\
3C~93.1  &  0.103   & 0.070 &40.2147  &40.0470  & \nodata & 3.806   \\
3C~129.0 & 0.12    & 0.087 & 38.0598  &37.9201  & \nodata   &0.243    \\
3C~132.0 &  0.079   & 0.081 &39.9738  &39.9847  &  LEG & 9.77   \\
3C~136.1 & 0.205   & 0.123   & 39.2786  & 39.0567 & \nodata   &2.64   \\
3C~180.0 &  0.228   & 0.488 &40.4605  &40.7910  & HEG    &16.78     \\
3C~196.1 &  0.149   & 0.136 &40.1756  &40.1360  & LEG   &8.72    \\
3C~197.1 &  0.121   & 0.091 &39.6925  & 39.5687 &   HEG &3.04   \\
3C~219.0 &  0.148   & 0.143  &40.0509  &40.0360  & BLRG  &1.90  \\
3C~227.0 &  1.931   & 0.569  &40.5184  &39.9878  & HEG$^{\mathrm{~c}}$ &21.76  \\
3C~234.0 &  0.941   & 2.050  &40.9068  &41.2449  &    BLRG &6.47  \\
3C~270.0 & 1.91  & 2.10 & 38.3014  &38.3425  & \nodata  &  0.63  \\
3C~285.0 & 0.433  &  0.711  &39.7925  & 40.0079   & HEG  &6.49  \\
3C~314.1 &  0.016   & 0.070 &38.7325  & 39.6781 & \nodata  & 3.13 \\
3C~319.0 &  0.027   & 0.016  &39.4047  & 39.4820 &  LEG &6.79 \\
3C~388.0 &  0.056   & 0.366  &39.0220  &40.1419  &  LEG &1.98  \\
3C~390.3 &  11.5    & 1.94  & 40.9081 &40.4398  & BLRG & 2.97 \cr
  \enddata
  \tablecomments{
    (1) 3CR source name;
    (2) measured H$\alpha$ (6563 \AA) line flux in erg s$^{-1}$ cm$^{-2}$;
    (3) total measured [O{\sc iii}] (5007 \AA) line flux in erg s$^{-1}$ cm$^{-2}$;
    (4) total H$\alpha$ luminosity in erg s$^{-1}$ and
    (5) total [O{\sc iii}] luminosity in erg s$^{-1}$ (fluxes converted to luminosity using $H_0 =  71$ km s$^{-1}$ Mpc$^{-1}$,
$\Omega_M = 0.27$, and $\Omega_{\Lambda} = 0.73$);
    (6) Source classification of host (FR~I = Fanaroff-Riley class I radio galaxy, QSO = quasar, BLRG = broad line radio galaxy, HEG = high-excitation FR~II galaxy, LEG = low-excitation FR~II galaxy, per the conventions in \citealt{jackson97});
    (7) largest measured angular size of line-emitting region (noted emission line 
in parentheses note the line for which the observed NLR appears largest, both 
lines will be listed if the NLRs cover approximately the same physical extent in 
both lines).
  }
  \tablerefs{(a) \citet{jackson97}, (b) \citet{buttiglione09}, (c) \citet{prieto93}}
\label{tab:tab2}
\end{deluxetable*}

\subsection{Image Alignment}

To facilitate  continuum subtraction it  was first necessary  to align
the  continuum exposures  with their  corresponding  drizzled emission
line images to  ensure registration down to a fraction  of a pixel.  A
first pass registration was made  by the world coordinate system (WCS)
to achieve rough  alignment of the continuum image  with its companion
line  emission data.   Subsequently, fractional  shifts  and rotations
were  manually applied  using  foreground stars  and  features on  the
galaxy  as  alignment   aides  (with  the  added  help   of  the  task
\texttt{imalign}).   Initial  image   registration  was  already  well
established  for  8  of  the  19  galaxies  in  our  sample  as  their
observations  were carried  out  using  the same  guide  stars in  all
(continuum,  \ha,  and  [O{\sc  iii}]) exposures.   The  remaining  11
observations included  small slews  of {\it HST}  to allow  for proper
placement of the  galaxy image on the WFC chip  (necessary to image at
the desired  LRF wavelength).  In  these cases guide  star acquisition
was required between exposures.  Registration for these 11 targets was
therefore  not initially  ideal, and  great care  was taken  to ensure
proper alignment through manually applied shifts and rotations.  After
the appropriate shifts and rotations  were applied, the WCS systems on
the images were registered with respect to one another so as to enable
proper WCS alignment when viewing the images.

\subsection{Continuum subtraction}

Here we describe the steps taken  to produce \ha and [O{\sc iii}] line
emission  images  that  are  effectively free  from  contamination  by
continuum light  from the galaxy.  The aligned  images were multiplied
by their exposure  times to set pixel values  (in electrons per second
for drizzled ACS images) to  total electrons.  The continuum image was
then scaled  by a factor determined  for each image  on a case-by-case
basis.   Initially,  this  factor  was  roughly  estimated  using  the
\texttt{synphot}    synthetic    photometry   package    (specifically
\texttt{calcphot}) in IRAF.  The  \texttt{calcphot} models served as a
qualitative guide  in roughly  establishing the scaling  factor, which
was later refined  iteratively through a trial and  error process that
involved  scaling and subtracting  the continuum  image from  the line
emission image,  and examining residual pixel values  on the resultant
image in  regions known to be  dominated by continuum  emission on the
original  unsubtracted data.   The  process was  repeated until  these
regions  possessed a  mode pixel  value very  close to  the background
level (which  was effectively  zero, after the  multidrizzle process).
Ultimately, the above strategies were employed in both qualitative and
quantitative  ways on a  case-by-case basis  for each  image, ensuring
that  the final subtracted  image could  be confidently  considered as
``pure'' line  emission within the  error established by  the absolute
photometry for  this dataset.  In Fig.~\ref{fig:contsub}  we provide a
representaive ``before and after'' comparison in the case of continuum
subtraction for 3C~33.

\begin{figure*}
\plottwo{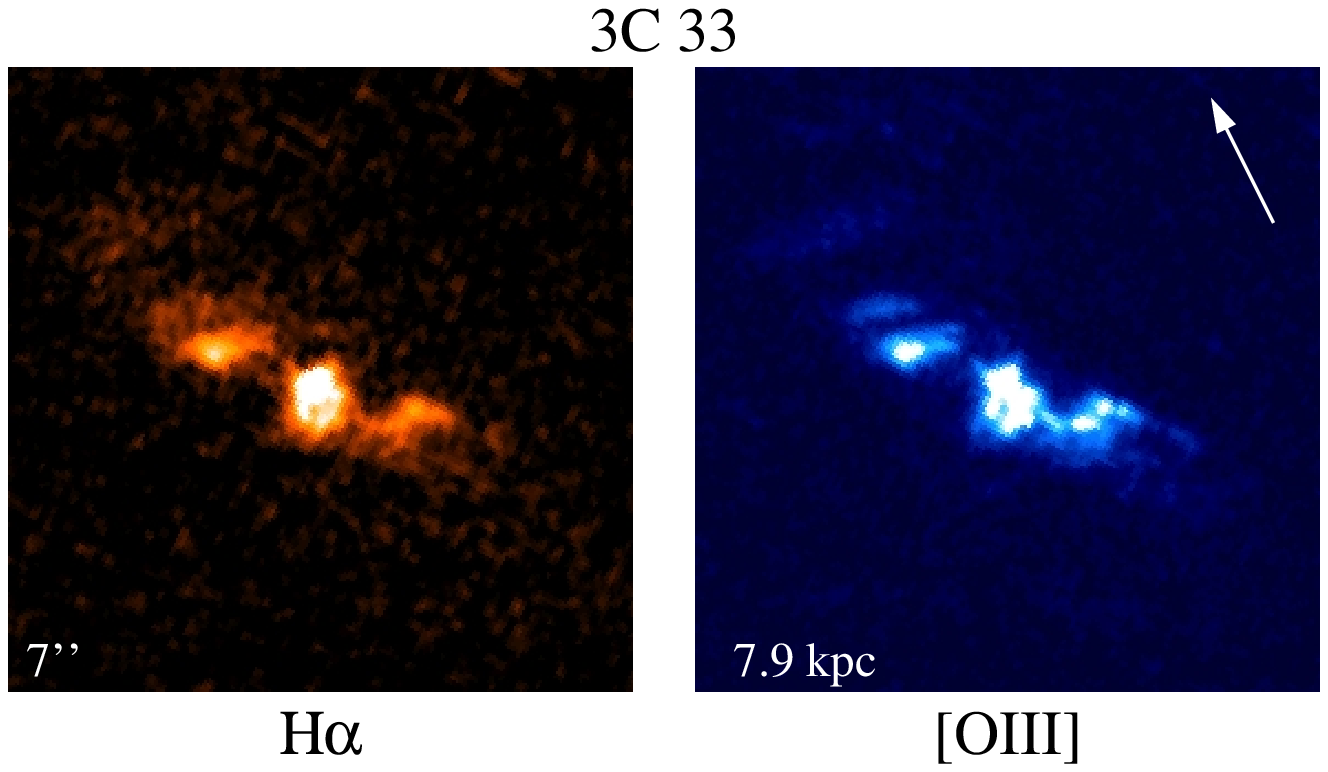}{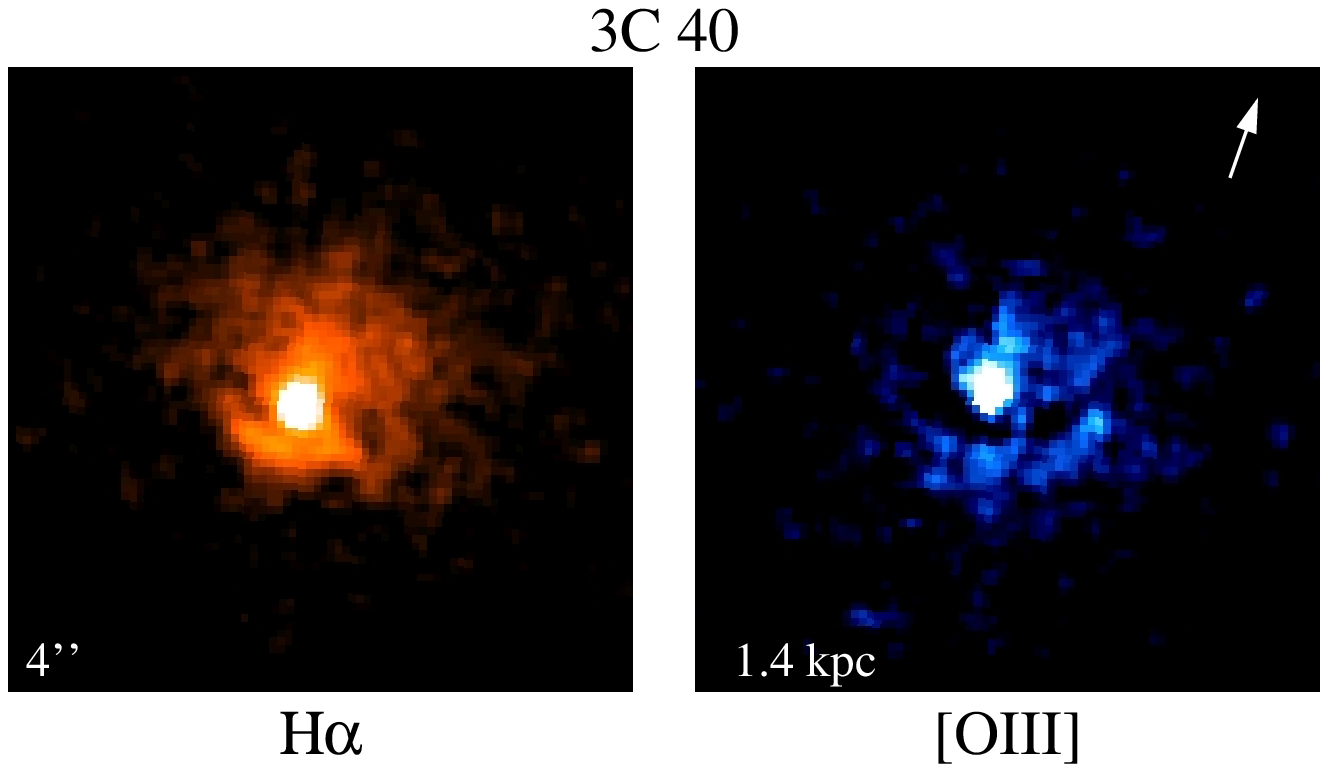}
\plottwo{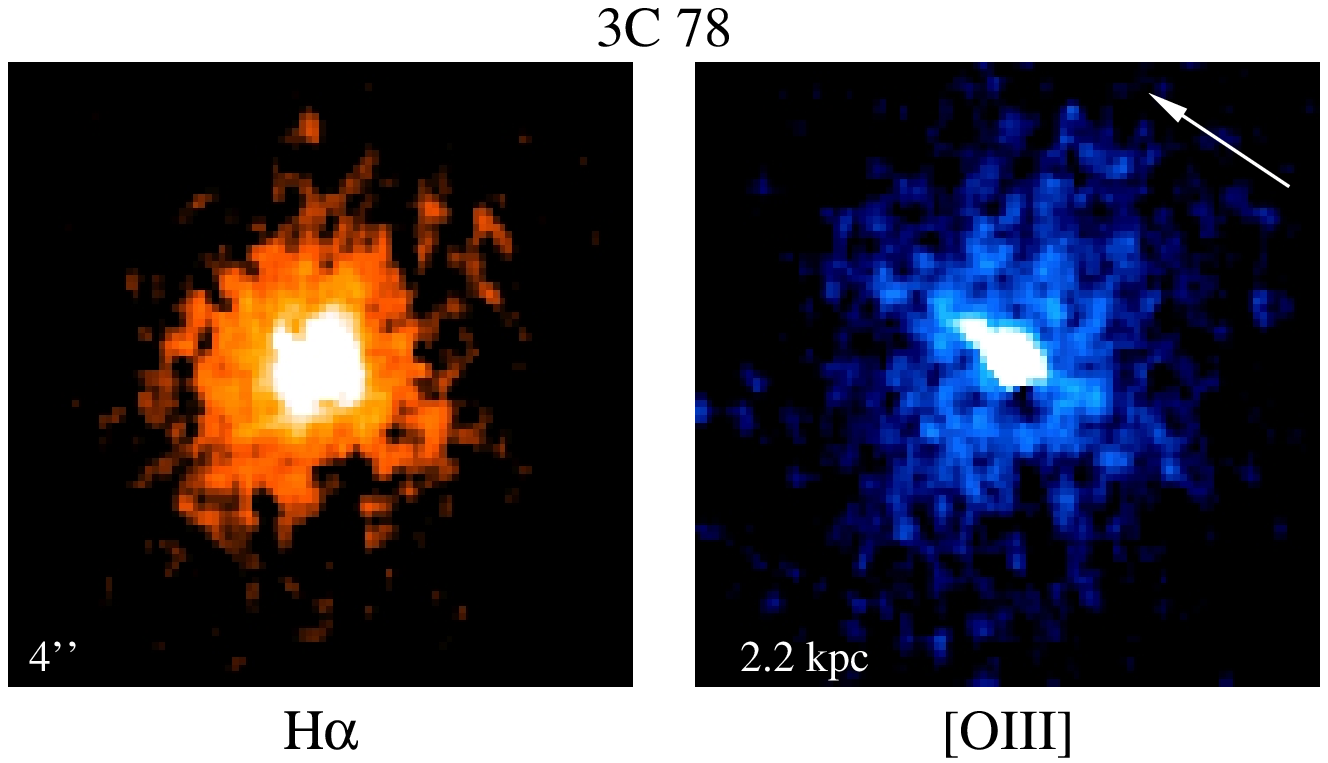}{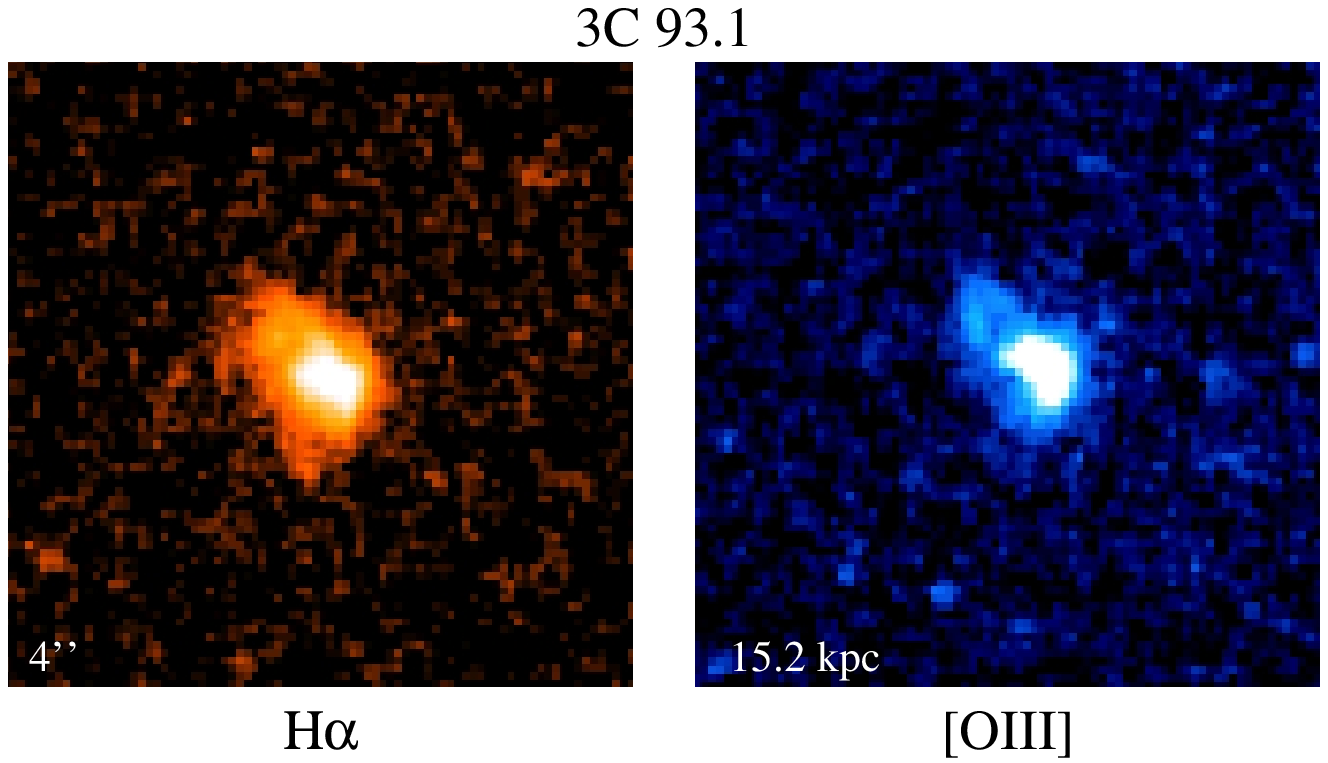}
\plottwo{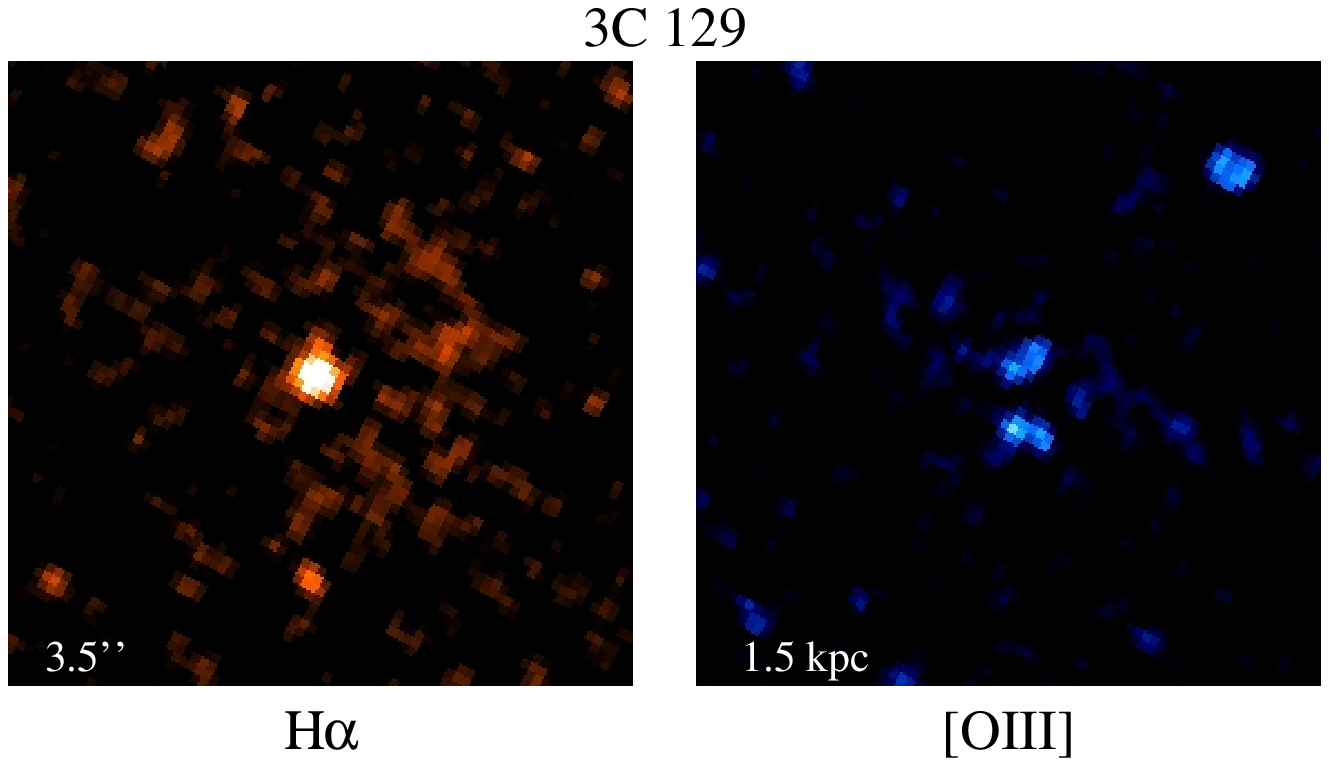}{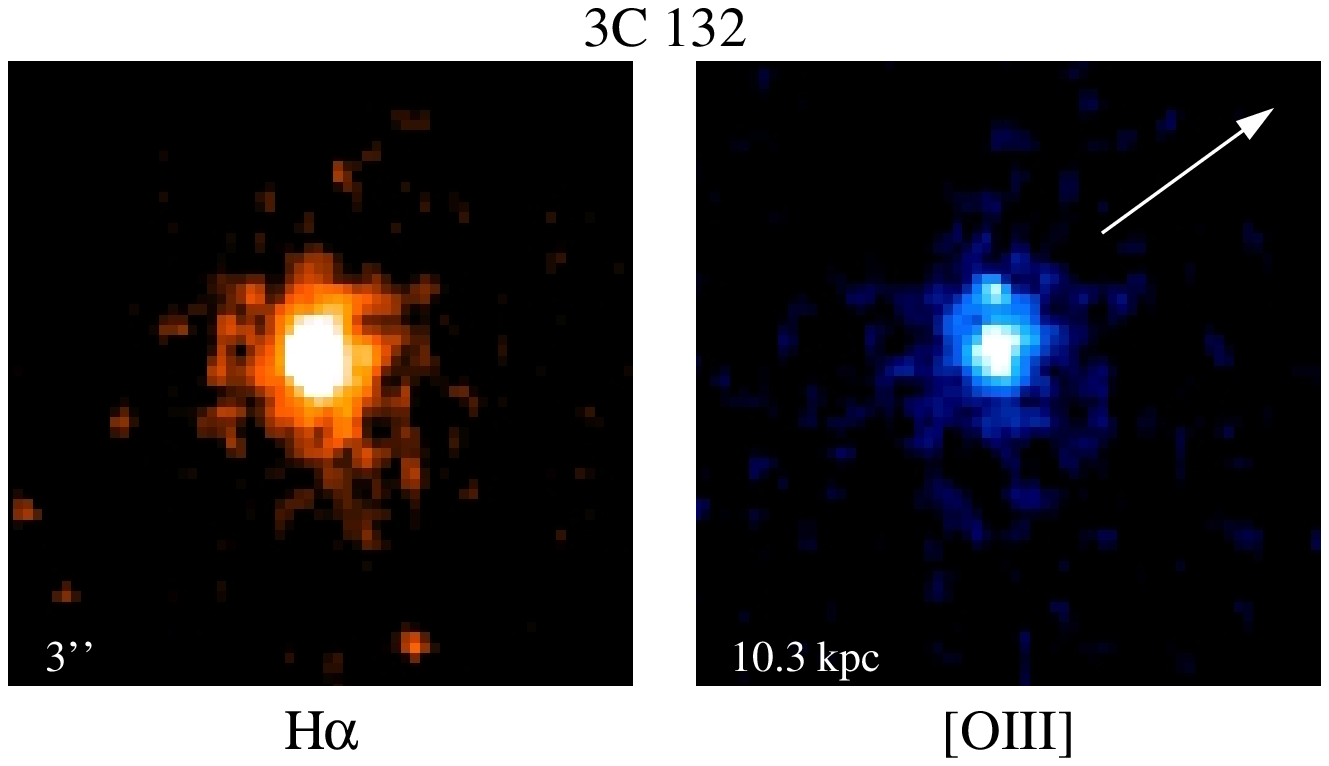}
\caption{ {\it HST}/ACS narrow-band images  of the \ha (left panel, in
  red) and [O{\sc  iii}] (right panel, in blue)  line emitting regions
  in our sample  of radio galaxies from the  3CR catalog. Contribution
  from  continuum has been  subtracted using  a strategy  discussed in
  \S3. In the bottom left corner of the left (right) panel we indicate
  the  size of  the field  of view  in arcsec  (kpc). Both  images are
  aligned, with East left and North  up. The arrows in the upper right
  corner  of some panels  indicates the  projected orientation  of the
  radio jet axis on the  sky, estimated from high resolution VLA radio
  maps (where  available). Galaxies are  listed in ascending  order by
  3CR  name.  See  Table  \ref{tab:tab1}  for more  details  on  these
  observations, and \S4 for descriptions of the emission line features
  of each target.  }
\label{fig:postage1}
\end{figure*}

\subsection{Emission line luminosity measurements}

For each  \ha and [O{\sc iii}]  image we have measured  the total flux
from  all detected  line  emission in  the  galaxy's central  regions.
Emission-line  fluxes  were  measured  from  the  continuum-subtracted
images  using the  \texttt{apphot} routines  in IRAF.   We  adopted an
``all the  flux you see'' approach  when establishing the  size of the
aperture  with which  the photometric  measurement was  performed.  In
this  way, we measure  ``total'' flux  as opposed  to flux  through an
aperture that is the same across the the entire dataset.

For consistency, we defined a ``source region'' by heavily smoothing a
copy of  each emission  line image  with a Gaussian  kernel set  to an
arbitrarily large  sigma.  The aperture  for the measurement  was then
set  to  the radius  corresponding  to  a  reasonable contour  in  the
smoothed image  enclosing all detected line emission  from the galaxy.
We  also defined  an  outer boundary  beyond  which the  image may  be
non-monochromatic  (due   to  the  ramp   filters,  see  \S2   for  an
explanation).  The region between  the ``source region'' and the outer
boundary defined the area in which the \ha and [O{\sc iii}] background
level was measured.

With these regions defined using the smoothed image, the corresponding
regions on  the original continuum-subtracted  images were identified.
The image  was then  examined to establish  the standard  deviation of
background, and the pixel values  were thresholded to a value of order
one  sigma above the  background level,  corresponding to  $\sim 0.01$
electrons  per second.   Counts above  this threshold  were considered
signal and were  counted in the photometric summation  of flux values,
while  pixels below  this threshold  were read  as zero  and  were not
counted.

We  summed counts above  the pre-defined  threshold within  the source
region  (aperture),  and  then  subtracted  the  estimated  background
measured  similarly within  the intermediate  region (exterior  to the
source  region  but  interior  to the  outer  boundary).   Photometric
conversion  was then  applied by  scaling the  residual  (source minus
background)  value  by the  inverse  sensitivity  (the {\sc  photflam}
keyword), converting the value from electrons per second to flux units
in erg cm$^{-2}$ s$^{-1}$  \AA$^{-1}$.  Flux measurements were made on
unrotated images  to avoid pixel interpolation  errors.  Emission line
flux  values  were  converted   to  luminosities  via  the  luminosity
distance.  Non-rotated  images  were  used  for  flux  calibration  to
minimize  pixel interpolation  errors.   Ultimately, the  calibrations
allow us  to measure emission line  flux with an  uncertainty of $\sim
15\%$.

\section{Results \& Discussion}

In  Figs.~\ref{fig:postage1}  -   \ref{fig:postage3}  we  present  the
continuum-free  line emission  images listed  by ascending  3C number.
\ha  images are  shown to  the left  and are  presented in  orange and
    [O{\sc  iii}]  images are  shown  to  the  right in  blue.   Where
    possible we have estimated  the projected orientation of the radio
    jet axis  on the  sky from high  spatial resolution  archival Very
    Large  Array  (VLA)  radio   maps  retrieved  from  the  NASA/IPAC
    Extragalactic                Database               (NED\footnote{
      \texttt{http://nedwww.ipac.caltech.edu/}}),   indicated   by  an
    arrow in the top right  corner of the galaxy's [O{\sc iii}] image.
    This  orientation  was {\it  qualitatively  }  estimated from  the
    apparent inner  jet axis  for FR~I radio  galaxies, and  along the
    hotspot-to-hotpsot  axis  for  FR~II  radio galaxies.   See  Table
    \ref{tab:tab3} for  a summary of the radio  properties for objects
    in the sample.  In instances  where the figure does not display an
    estimate  for  the radio  jet  orientation,  data  was either  not
    readily  available,  was  of  insufficient resolution  to  make  a
    confident estimate,  or was cospatial with the  source on galactic
    scales (i.e.  morphologically similar to CSS sources).

\subsection{Description of individual sources}

\begin{figure*}
\plottwo{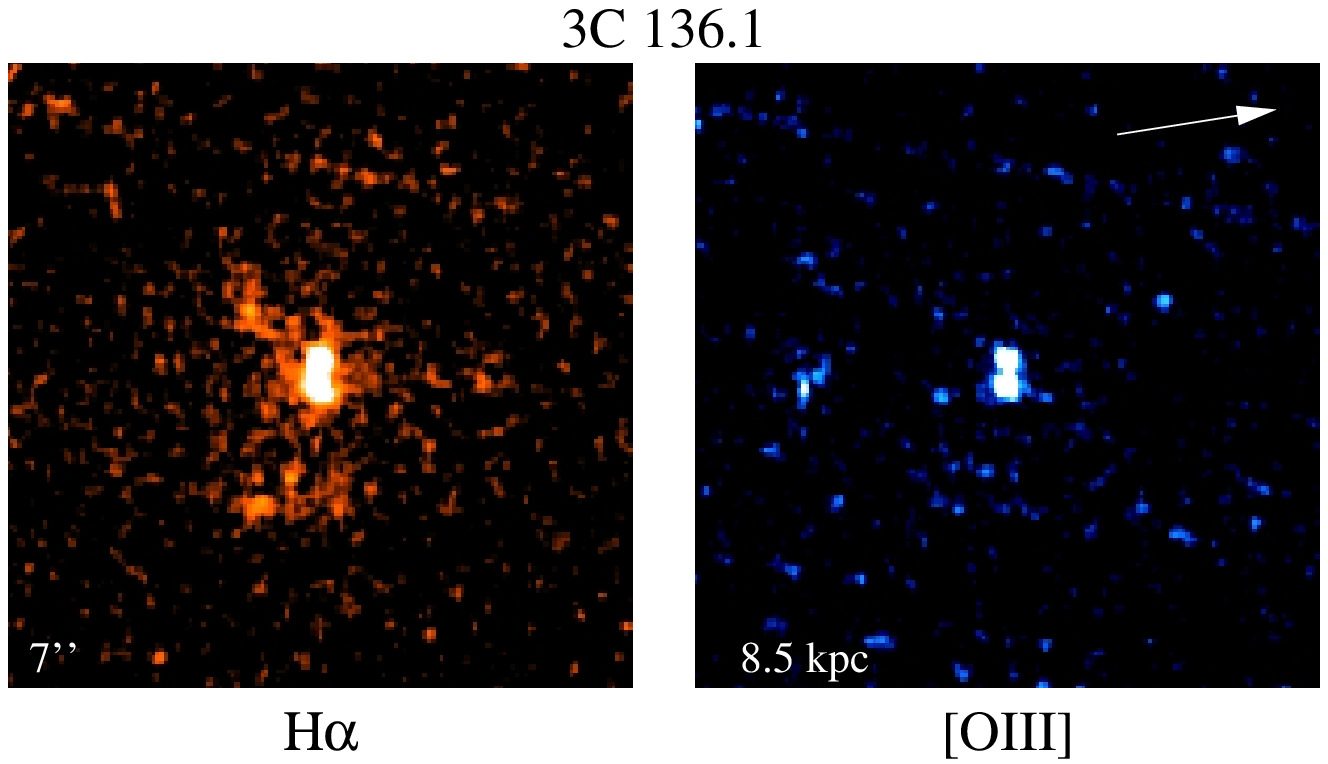}{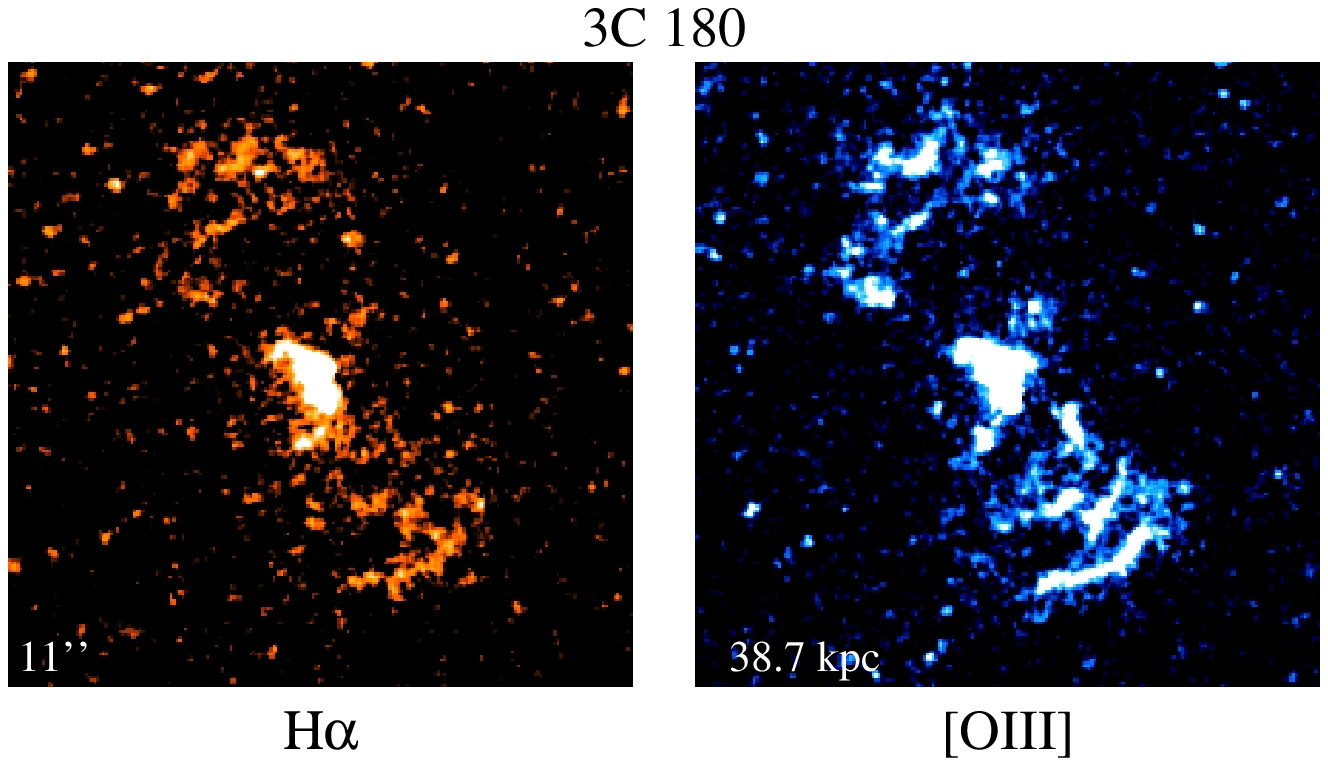}
\plottwo{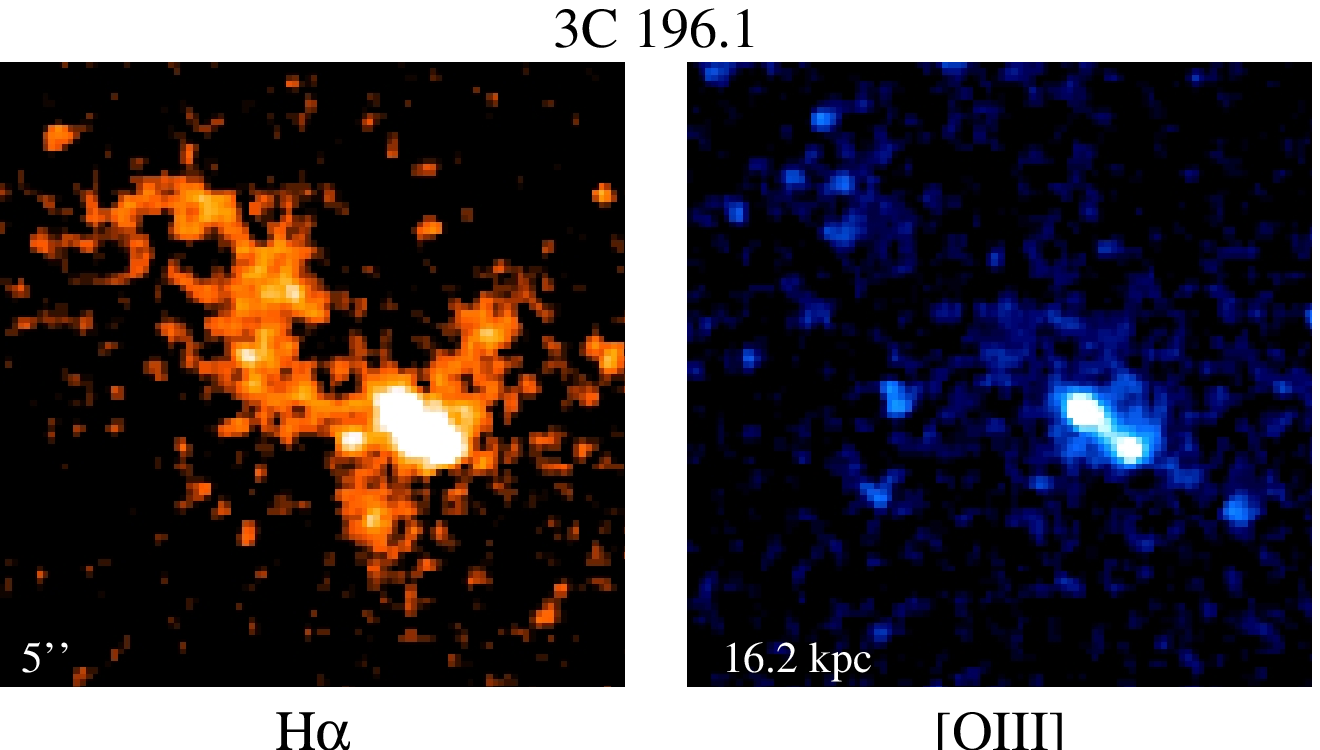}{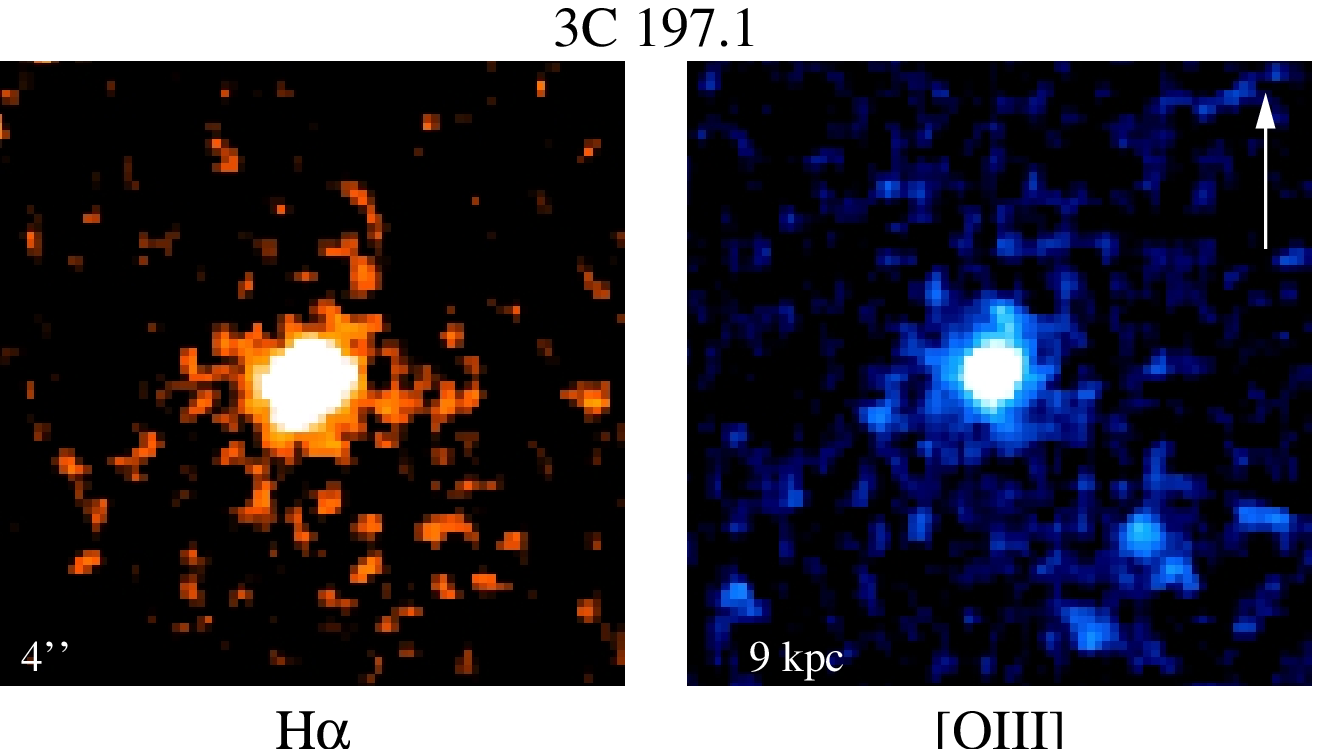}
\plottwo{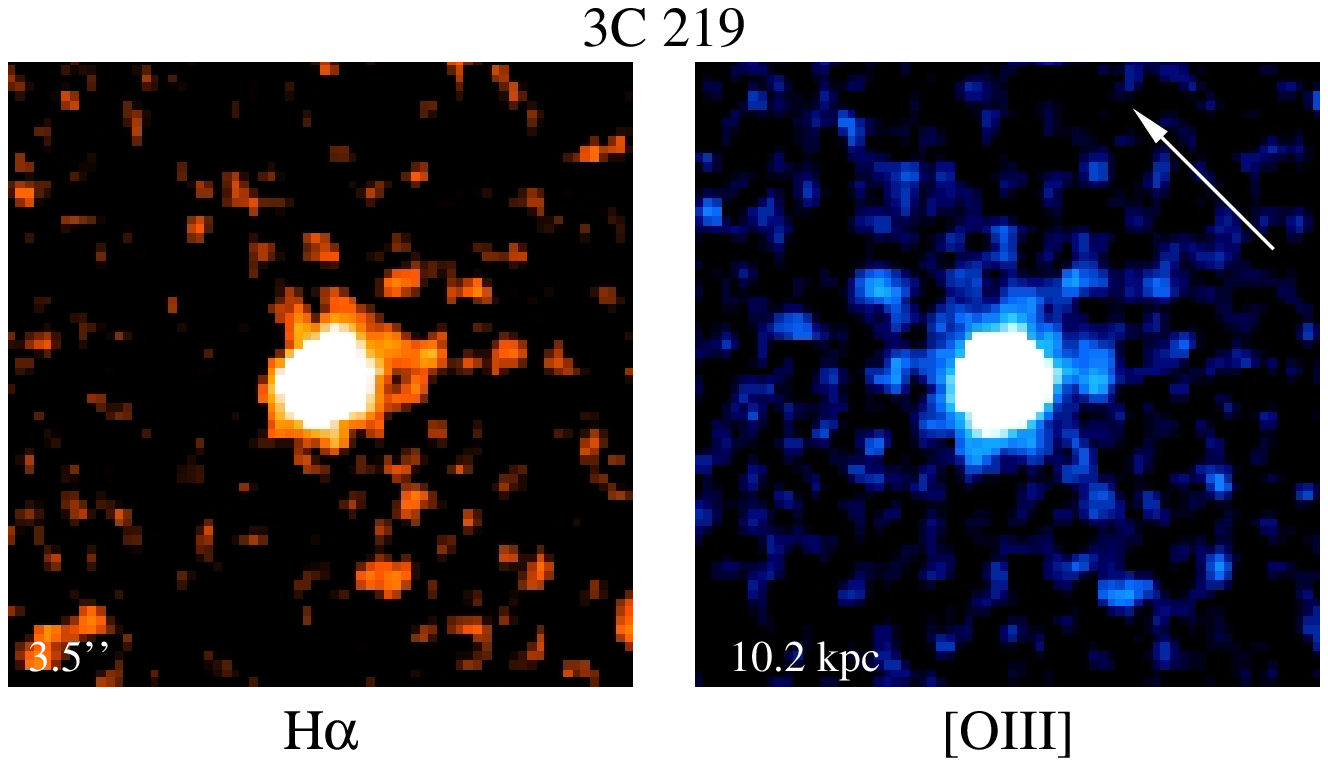}{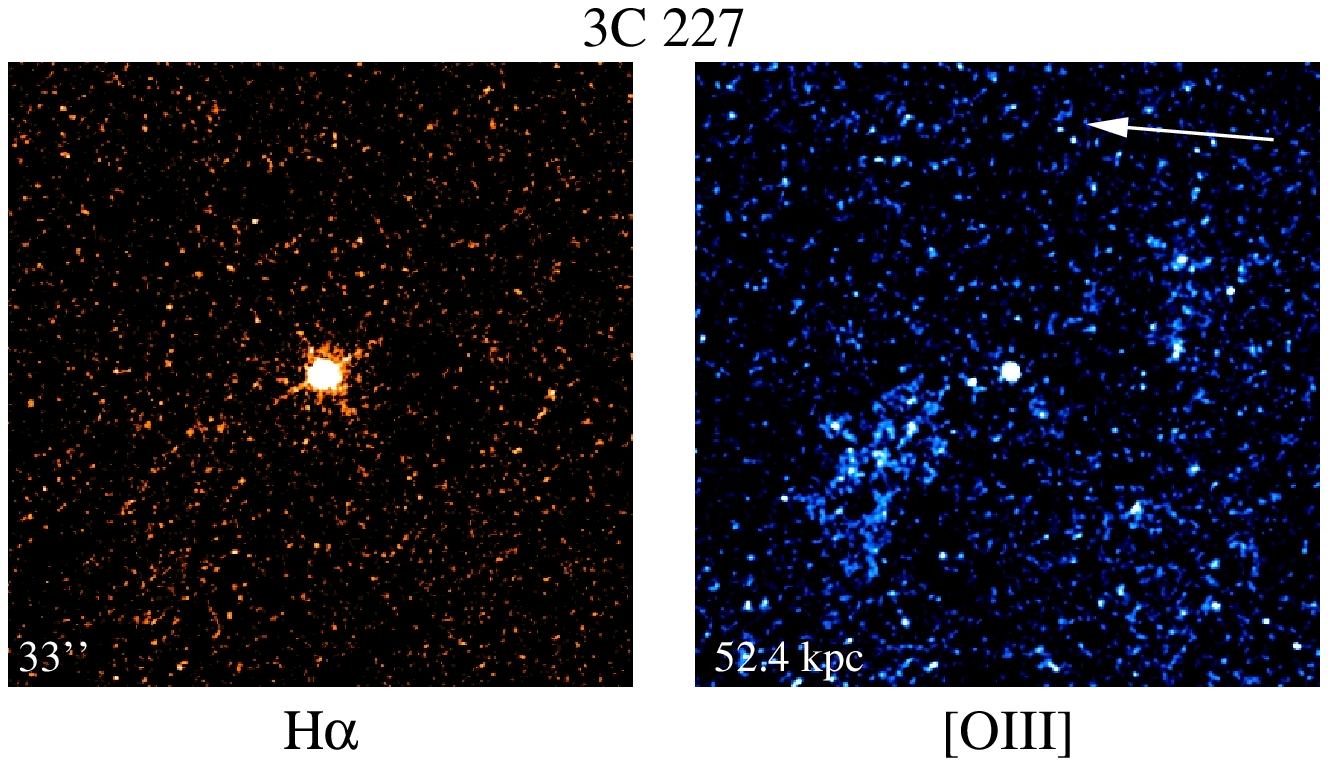}
\caption{Same as  Fig.~\ref{fig:postage1}. East is left,  North is up.
  Note that panels labeled ``H$\alpha$'' actually show \ha emission.}
\label{fig:postage2}
\end{figure*}

\begin{figure*}
\plottwo{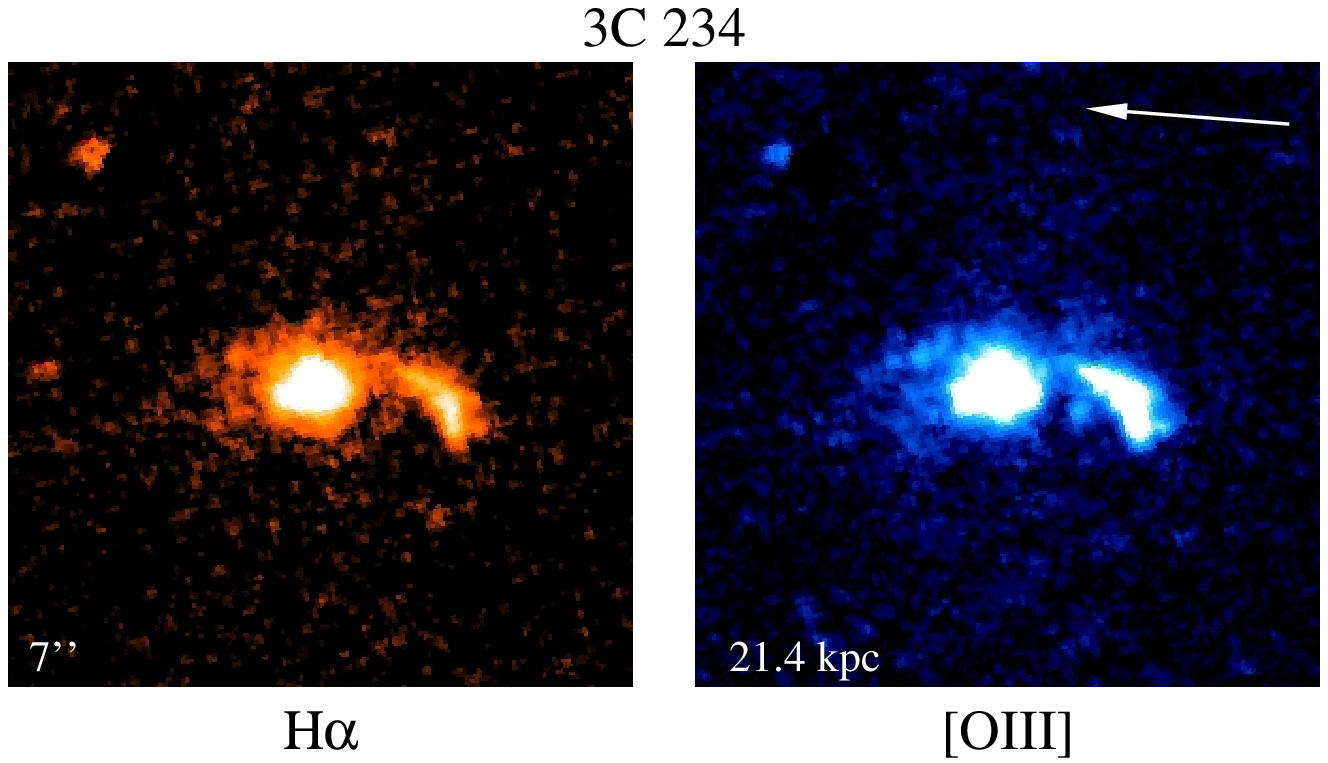}{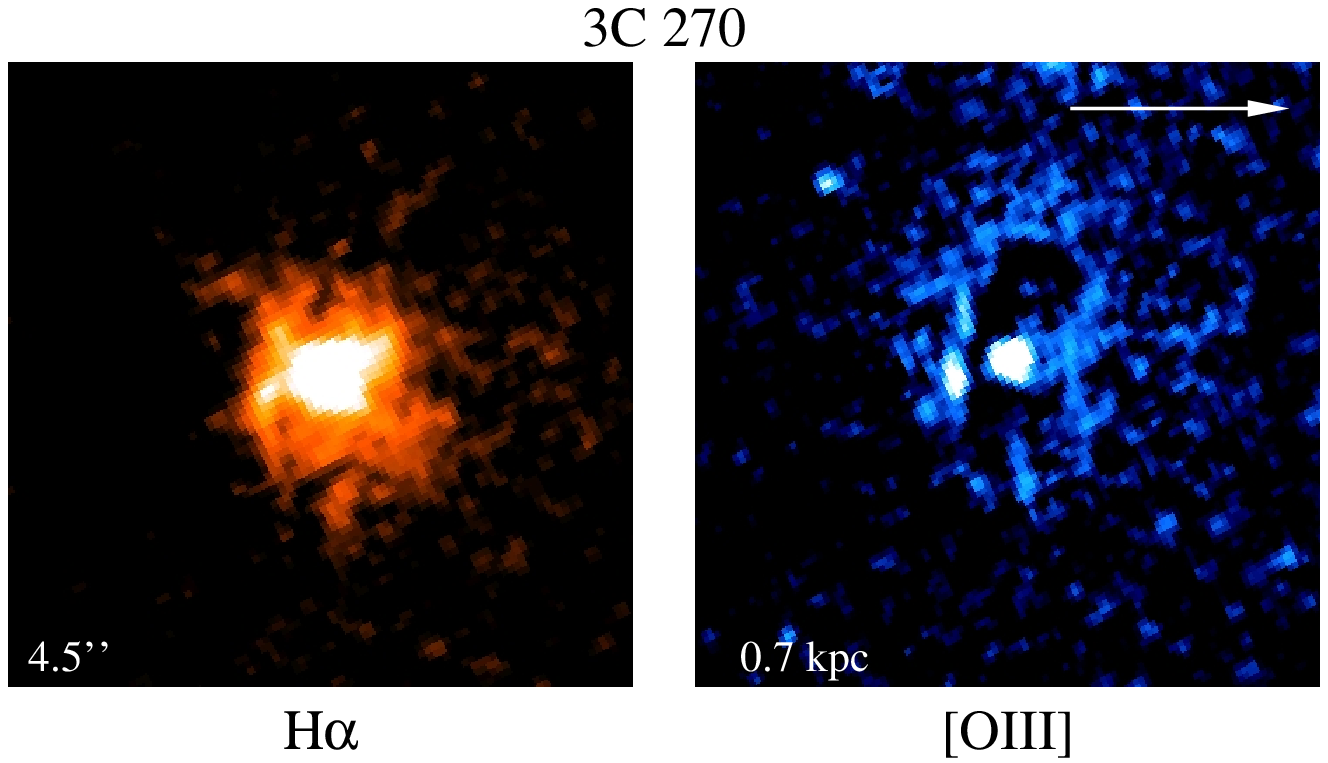}
\plottwo{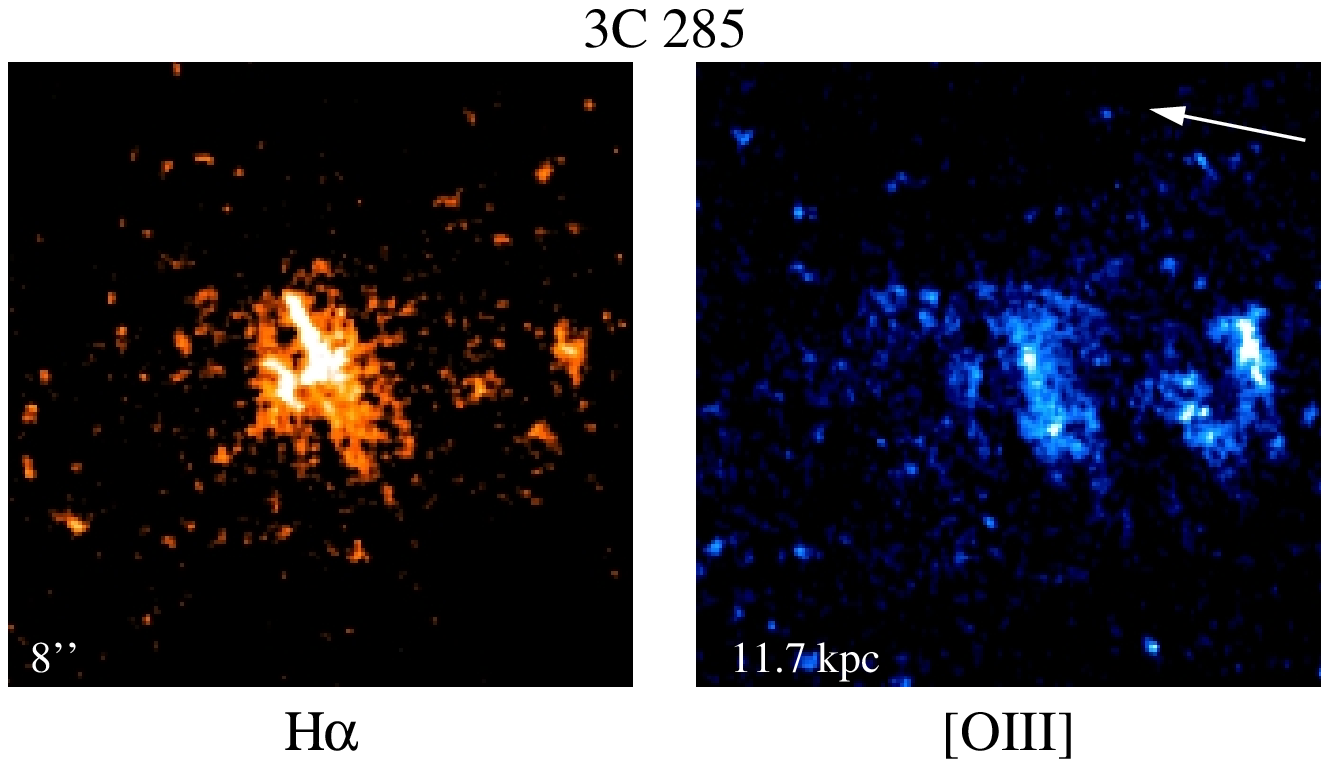}{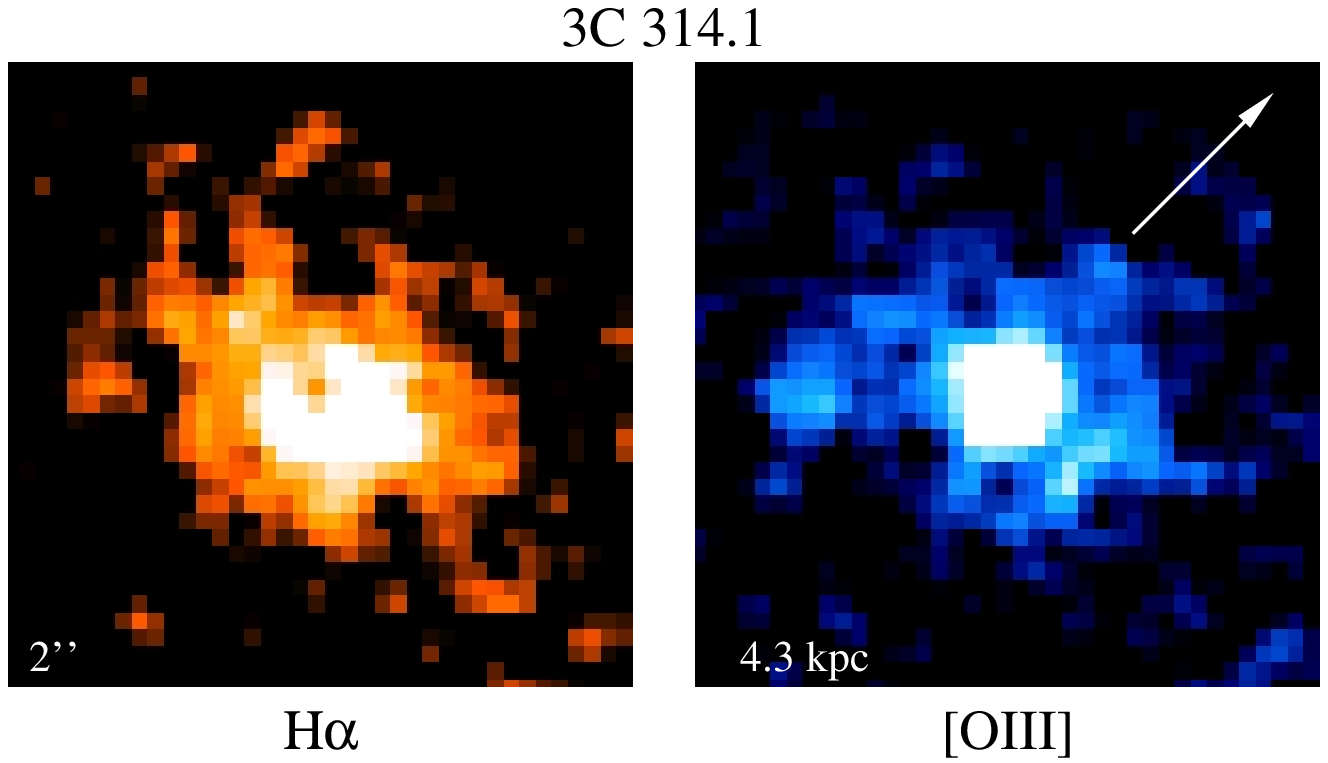}
\plottwo{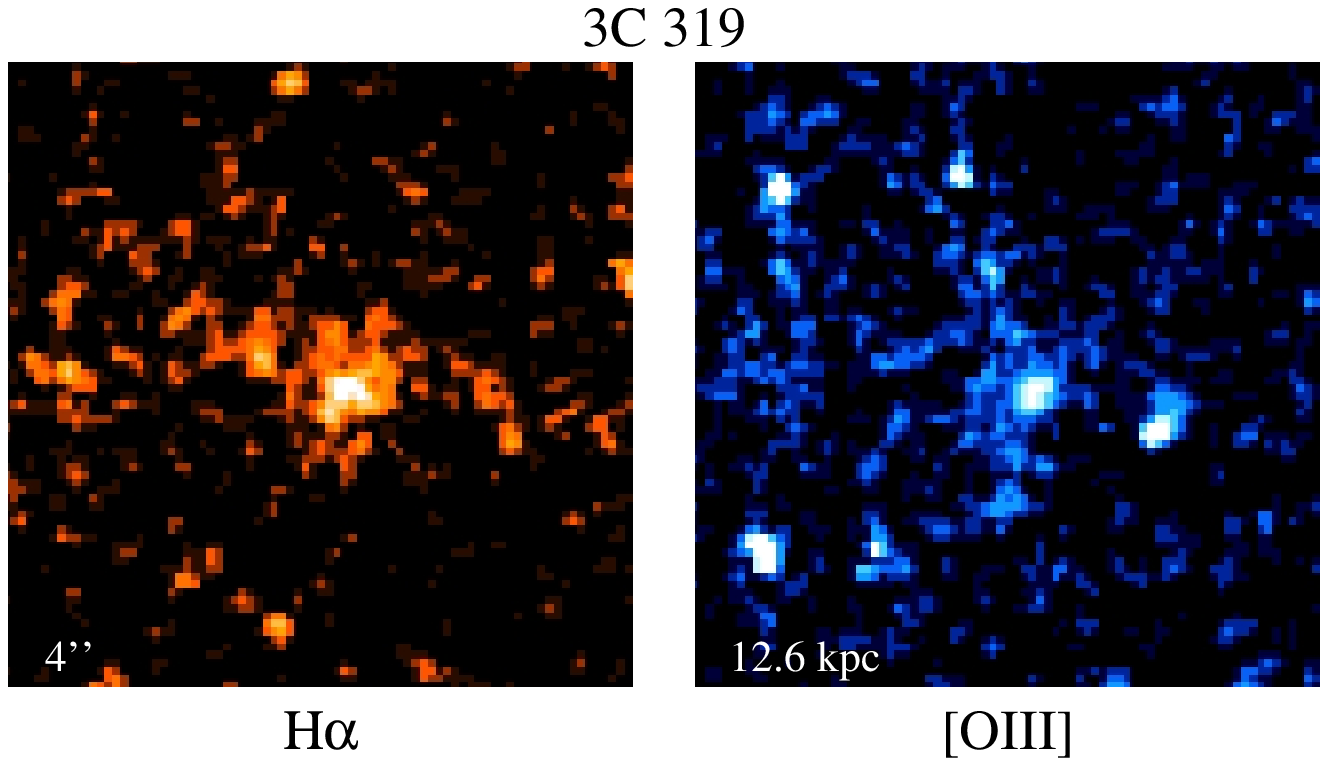}{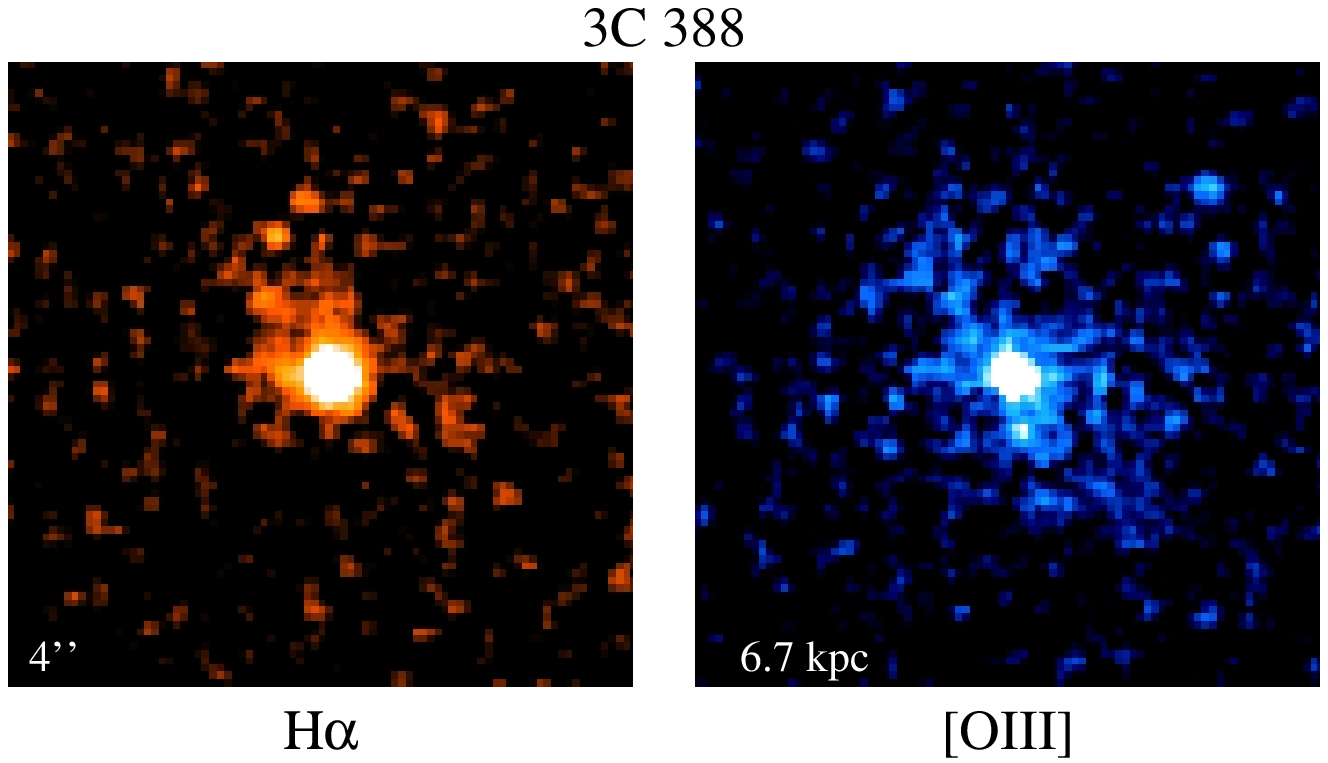}
\plottwo{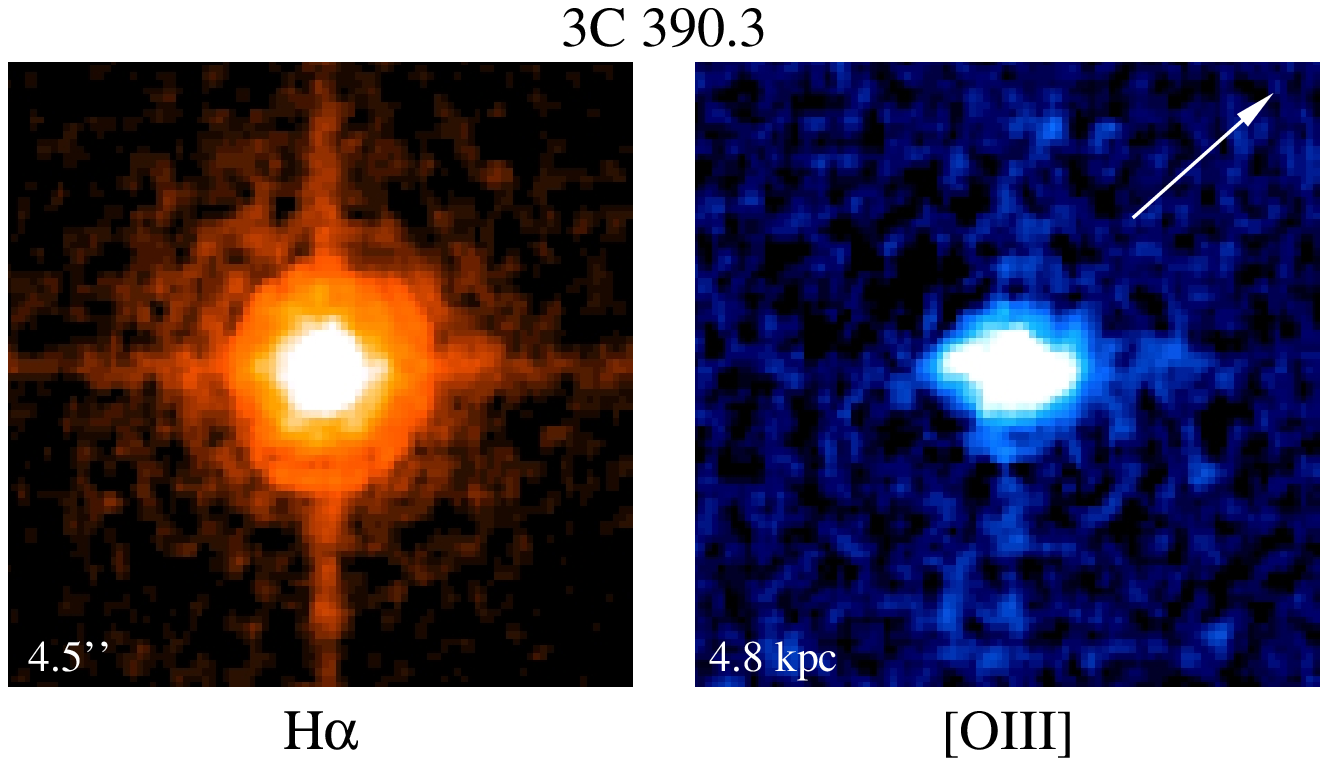}{blank.eps}
\caption{Same as Fig.~\ref{fig:postage1}. East is left, North is up.
 Note that panels labeled ``H$\alpha$'' actually show \ha emission.}
\label{fig:postage3}
\end{figure*}

In the  subsections below  we discuss the  observed morphology  of the
high surface brightness \ha and [O{\sc iii}] line emission detected in
each galaxy.  Objects are listed in ascending order by the name of the
source.  For information  relating to the observations of  each of the
below  galaxies,  see  Table  \ref{tab:tab1}.  For  quantitative  data
regarding the optical and radio  properties of each object, see Tables
\ref{tab:tab2} and \ref{tab:tab3}, respectively.  All images discussed
below are  presented in Figs.  \ref{fig:postage1}, \ref{fig:postage2},
and \ref{fig:postage3}.

\subsubsection{3C 33; $z= 0.059$} 
Extended regions of high surface  brightness line emission are seen in
both the  \ha and [O{\sc  iii}] images.  The low-  and high-excitation
regions trace a general ``integral sign'' shape extending $\sim 5$ kpc
and oriented  NE to SW.  In  both images a fainter,  detached shell of
low-  and  high-excitation  emission   is  observed  NE  of  the  main
``integral sign''  feature.  The [O{\sc iii}]  features more obviously
clumpy,  brighter hotspots  along the  lane of  emission and  near the
nucleus  than  does the  \ha  image.  Long  slit spectroscopy  of  the
extended    line   emission    obtained   by    \citet{simkin79}   and
\citet{heckman85}  suggested that  the line  emitting gas  is rotating
along an axis oriented at  a position angle (P.A.)  of $\sim 19^\circ$
with respect to the FR~II radio jet axis. (Fig.~\ref{fig:postage1}).

\subsubsection{3C  40; $z=  0.017$}
A dusty  disk on 100 pc  scales is seen  in both \ha and  [O{\sc iii}]
images, marked  by the absence  of emission along a  circumnuclear rim
that is  more prominent in the [O{\sc  iii}] image than it  is in \ha.
Low- and  high-excitation line emission  is seen tracing the  edges of
the  disk on the  southern side,  while [O{\sc  iii}] emission  on the
northern side  of the disk appears less  extensively distributed.  The
band marking the  disk is not continuous, rather in  both images it is
interrupted   by  low-surface   brightness   line-emission  apparently
connecting the  southern and  northern sides of  the disk in  the same
general  location.  Note  that the  apparent  major axis  of the  disk
appears roughly orthogonally oriented  with respect to the FR~II radio
jet axis.  (Fig.~\ref{fig:postage1}).

\subsubsection{3C 78; $z= 0.028$} An  optical jet is
seen in both \ha and [O{\sc  iii}] emission to extend from the core in
a  northeasterly  direction  (projected  on  the  sky).   The  bright,
unresolved  nucleus  in  both  images  is surrounded  by  a  generally
isotropic region  of narrow-line emitting  gas in both \ha  and [O{\sc
    iii}].   \citet{sparks00} noted  the presence  of a  face-on dusty
nuclear  disk in this  FR~I radio  galaxy.  Both  distributions appear
slightly  more extended  in  the plane  perpendicular  to the  optical
synchrotron   radio   jet    axis   described   by   \citet{sparks95}.
\citet{perlman06}  notes  the  presence   of  an  emission  line  knot
cospatial  with the  radio jet  axis. That  we apparently  detect both
high-  and  low-excitation emission  along  the  jet  axis may  be  an
important    result,    and     is    worthy    of    future    study.
(Fig.~\ref{fig:postage1}).

\subsubsection{3C  93.1;  $z=  0.244$} A compact distribution of 
narrow-line emitting gas is  seen surrounding the nucleus and extended
$\sim  3.8$ kpc  primarily towards  the NE.   The  lopsided morphology
observed in both the \ha and  [O{\sc iii}] emitting gas is evidence of
the   presence   of   an   anisotropic  nuclear   ionizing   radiation
field. 3C~93.1 is a CSS source. (Fig.~\ref{fig:postage1}).

\subsubsection{3C   129; $z=  0.021$} 
Narrow-line  emission  associated   with  3C~129  is  only  marginally
detected  in both  filter configurations.   Compact \ha  emission with
very  low surface  brightness is  observed  to be  cospatial with  the
nucleus. Conversely, the  nucleus is not detected in  the [O{\sc iii}]
image, where instead only very faint extended emission that appears to
be associated  with the  nucleus extends 1$\arcsec$  to the  north and
south. 3C~129  is one of  the four FR~I  radio galaxies in  our sample
(Fig.~\ref{fig:postage1}).

\subsubsection{3C  132; $z= 0.214$}   
A  bright nucleus  surrounded  by fainter,  more  diffuse emission  is
observed  in both  \ha and  [O{\sc iii}]  images. The  nucleus appears
fainter in [O{\sc iii}] and what  appears to be a faint ``hotspot'' is
located       directly      to       the       north      of       the
nucleus. (Fig.~\ref{fig:postage1}).

\subsubsection{ 3C 136.1; $z= 0.064$}  
In  \ha,  an  elongated  nucleus  is seen  with  major  axis  oriented
north-south on  the sky. Faint, disjoint  distributions of narrow-line
gas surround the nucleus in  two filamentary plumes extending from the
core along  a N-S  axis roughly perpendicular  to the FR~II  radio jet
axis.   Clumpy patches of  disparate, low-surface  brightness emission
are observed at radii of $\sim 1\arcsec$.  In [O{\sc iii}] there is an
apparent double  nucleus with hotspots more clearly  separated than in
the \ha image.  The north-south filaments seen in \ha are not detected
in [O{\sc iii}]. (Fig.~\ref{fig:postage2}).

\subsubsection{3C 180; $z= 0.220$} 
A  bi-conical series  of shells  is observed  in both  \ha  and [O{\sc
    iii}], reminiscent  of other bi-conical  NLR morphologies observed
in  some  Seyfert  II  galaxies  (e.g.,  Mrk 3,  Mrk  573,  NGC  5252,
\citealt{capetti99}).  At  both low- and high-excitation  we observe a
bright, lopsided emission region  cospatial with the nucleus.  Fainter
``plume-like''  features extend  from both  the northern  and southern
extremities of  the extended core of emission.   The bi-conical series
of shells of  emission extends $\sim 8$  kpc to both the NE  and SW of
the core. The abundance of optical line emitting gas along a preferred
axis suggests the presence  of a highly anisotropic ionizing radiation
field. 3C~180 lacks a Fanaroff-Riley classification, and we are unable
to characterize the  morphology of its radio source  in any meaningful
way  given the  available data.   \citet{mccarthy95} note  that 3C~180
resides in a cluster.  (Fig.~\ref{fig:postage2}).

\subsubsection{3C  196.1; $z=  0.198$}  
A high surface brightness lane  of \ha emission ``snakes'' NE from the
bright  elongated  core,  reminiscent  of a  ``tadpole''  shape.   \ha
emission  is also  observed  extending  from the  core  along an  axis
perpendicular  to the  axis  of elongation  of the  ``tadpole''-shaped
distribution  of ionized gas.   In the  [O{\sc iii}],  high excitation
emission  is observed  in two  localized hotspots  cospatial  with the
elongated bright  core seen in  \ha, and the ``tadpole''  structure is
not   detected,  nor   is   the  emission   along  the   perpendicular
axis. (Fig.~\ref{fig:postage2}).

\subsubsection{3C  197.1; $z=  0.13$}  
Bright, unresolved nuclei  are seen in both images,  surrounded on all
sides  by a fainter  distribution of  line-emitting gas.   The nucleus
appears    larger     in    \ha    than    it     does    in    [O{\sc
    iii}]. (Fig.~\ref{fig:postage2}).

\subsubsection{3C  219; $z=  0.174$}  
Compact emission is seen in  both bands, as is faint, diffuse emission
surrounding       the       unresolved       source       at       the
center. (Fig.~\ref{fig:postage2}).

\subsubsection{3C 227; $z=  0.086$} 
Compact emission cospatial  with the galaxy's nucleus is  seen at both
low- and  high- excitation.  In  [O{\sc iii}] we observe  an extremely
large arm of emission extending $\sim 20$ kpc, representing by far the
largest distribution  of line-emitting gas  in our sample.   The large
structure is oriented  along an axis roughly perpendicular  to that of
the  FR~II radio jet.  \citet{prieto93} studies  this object  in great
detail. (Fig.~\ref{fig:postage2}).

\subsubsection{3C 234; $z= 0.184$}  
The  prominent tidal arm  described by  \citet{carleton84} is  seen in
high surface brightness \ha and [O{\sc iii}] emission.  To the east is
the bright  quasar-like nucleus  of the galaxy,  and in both  images a
bright feature  is seen  extending 0.5$\arcsec$ from  its northwestern
edge. The  overall elongation of  the emission line  region, including
the tidal arm, is in  a projected orientation that is roughly parallel
with     that     of    the     radio     jet     axis    on     large
scales. (Fig.~\ref{fig:postage3}).

\subsubsection{3C 270; $z = 0.0077$}  
The $\sim$ 120 pc dusty  disk originally studied by \citet{jaffe93} is
notably absent of emission in  both \ha and [O{\sc iii}], though there
remains some  \ha emission along the  inner regions of  the disk.  The
disk is  largely edge-on with respect  to the line of  sight though is
inclined such that the western side of the disk ``faces'' the observer
slightly.   Note that  the western  half of  the galaxy  is noticeably
brighter than the eastern half.   Cones of high surface brightness \ha
emission  are seen  extending from  the unresolved  nucleus  from both
sides of  the disk, and are  elongated along the direction  of the jet
(east-west on the  sky, and nearly perpendicular to  the major axis of
the dusty disk).  Nuclear [O{\sc iii}] emission is also seen extending
from both sides of the disk,  though seems to be largely absent on the
disk itself. (Fig.~\ref{fig:postage3}).

\subsubsection{3C  285; $z=  0.079$}  
A complex, filamentary system of dust lanes bisects the nuclear region
of the  galaxy in a  northeasterly direction and perpendicular  to the
line of sight.  The extinction associated with the predominant lane in
this system is seen in the \ha, where we observe an extended region of
high surface  brightness emission with complex  morphology.  A bright,
dense column  of line-emission extends $\sim  1\arcsec$ northward from
the nucleus at  the center of the image.  Slightly  to the southeast a
small, bright  ``integral sign'' feature is seen  in general alignment
with the large dense column.  Both of these features are surrounded by
fainter  narrow-line emitting  gas arranged  in filamentary  plumes as
well  as clumpy  patches.   In the  [O{\sc  iii}] image  we observe  a
dramatically different  morphology than is  apparent from the  \ha.  A
high surface brightness lane  of emission ``snakes'' westward from the
center of the  galaxy to form a disjoint ``W''  shape.  The nucleus of
the galaxy, such as it is,  is not discernible from the image.  Bright
patches  of  emission  populate  the  prominent  filament  at  regular
intervals  along its length.   A region  of particularly  high surface
brightness     emission     is     noted    at     its     westernmost
extremity. (Fig.~\ref{fig:postage3}).

\subsubsection{3C  314.1;  $z=  0.119$}   
Diffuse \ha emission is distributed in a complex morphology, while the
[O{\sc iii}]  emission is decidedly  more compact, albeit  with faint,
wispy  plumes  extending from  the  nucleus  and  slightly toward  the
southwest. (Fig.~\ref{fig:postage3}).

\subsubsection{3C  319; $z= 0.192$}  
Very faint line emission is  detected in both images, distributed in a
diffuse,   complex   morphology  at   extended   distances  from   the
nucleus.  The [O{\sc  iii}]  is  decidedly more  compact  than is  the
\ha. (Fig.~\ref{fig:postage3}).

\subsubsection{3C   388;  $z=  0.092$}   
A  bright unresolved  nucleus  is  seen in  both  images.  Small,  low
surface  brightness   ``plumes''  can  be  seen   extending  toward  a
northeasterly  direction in the  \ha, while  a more  prominent conical
feature is seen on the southwestern  side of the nucleus in the [O{\sc
    iii}] image. (Fig.~\ref{fig:postage3}).

\subsubsection{3C  390.3; $z= 0.056$} 
Very  bright \ha  emission  dominates the  image  from the  unresolved
nucleus.  The noticeably rhomboidal shape of the nucleus in the [O{\sc
    iii}] is  likely a result of  the diffraction spikes  seen in both
images. (Fig.~\ref{fig:postage3}).

\begin{figure*}
\plottwo{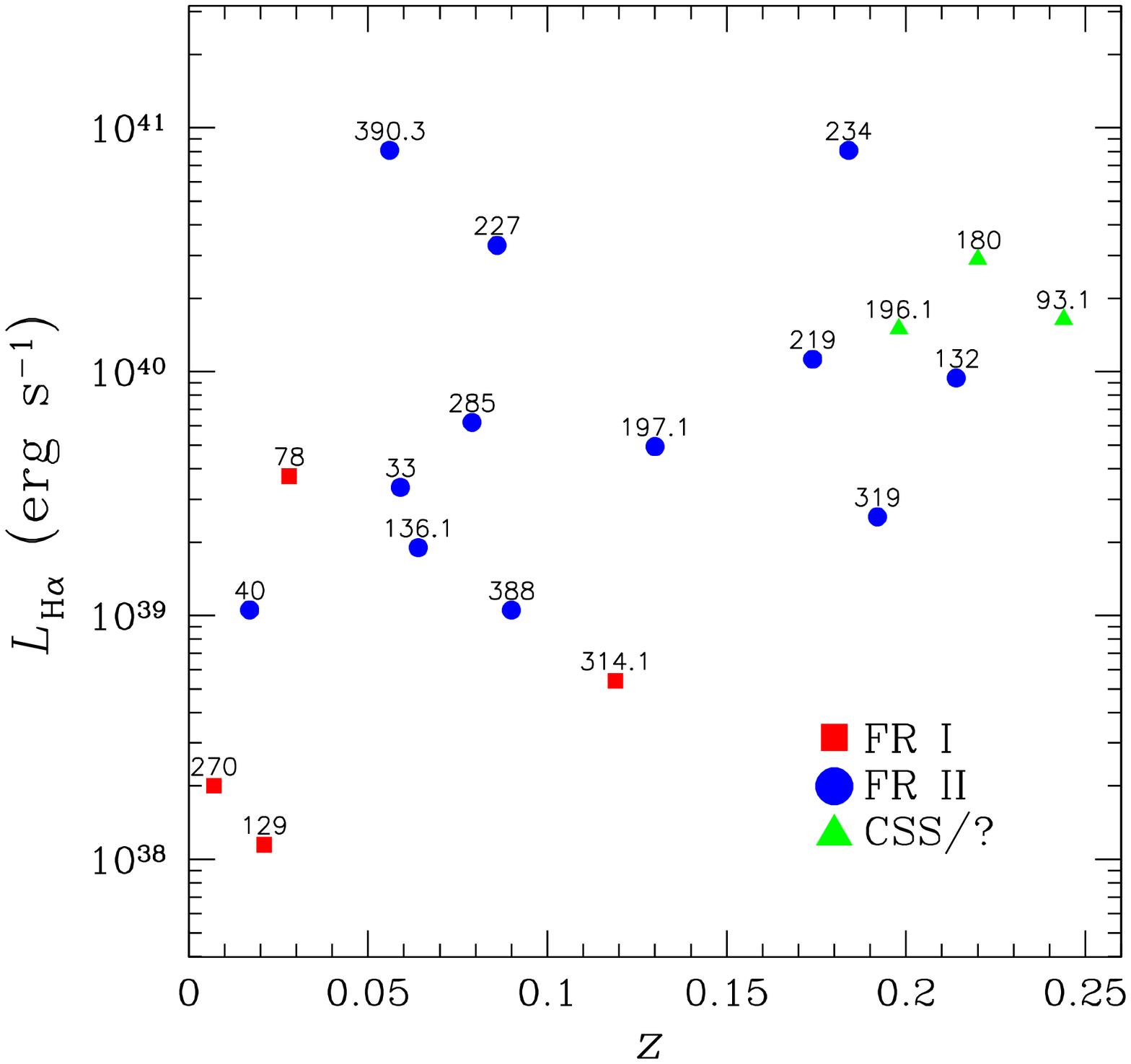}{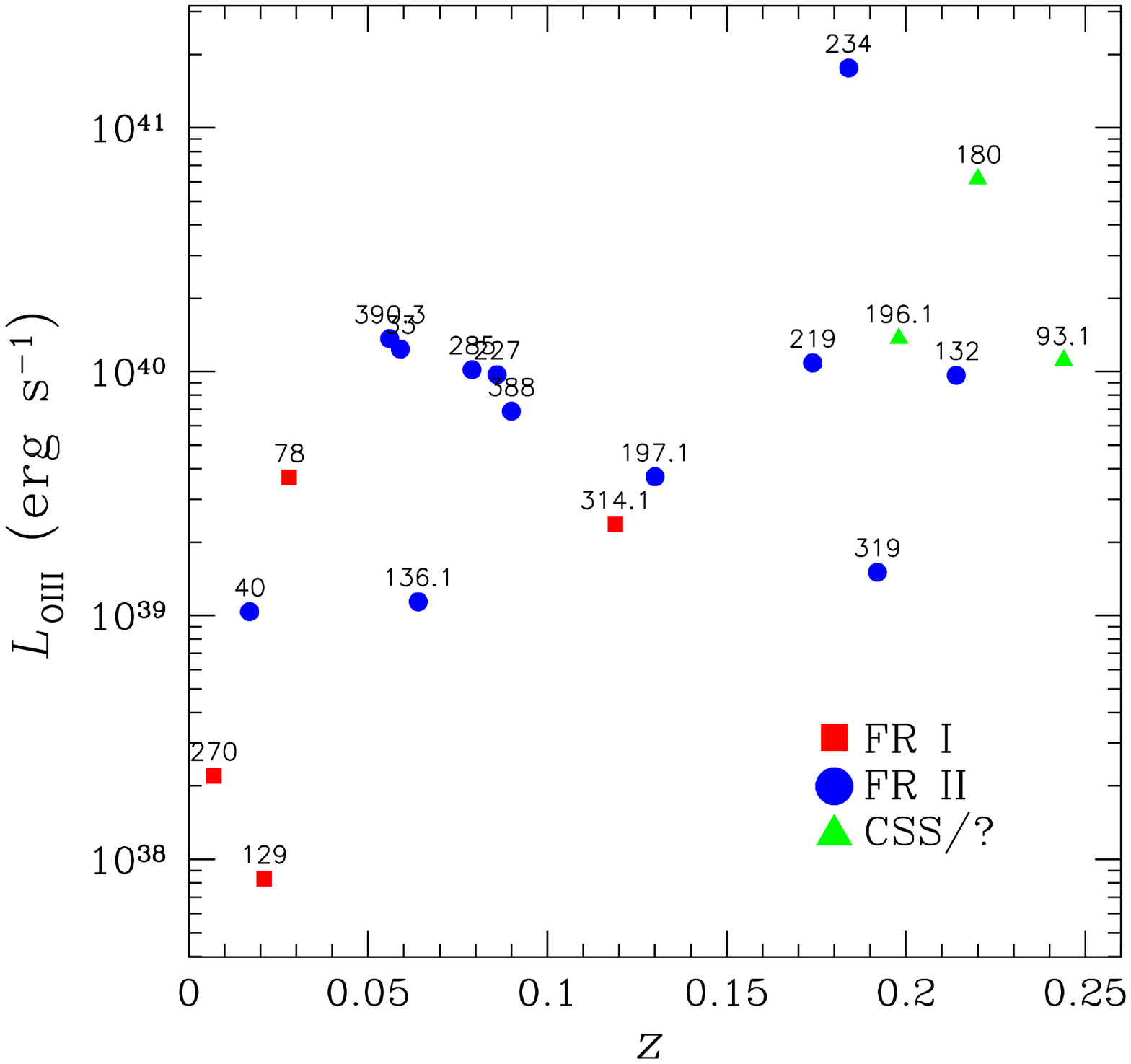}
\plottwo{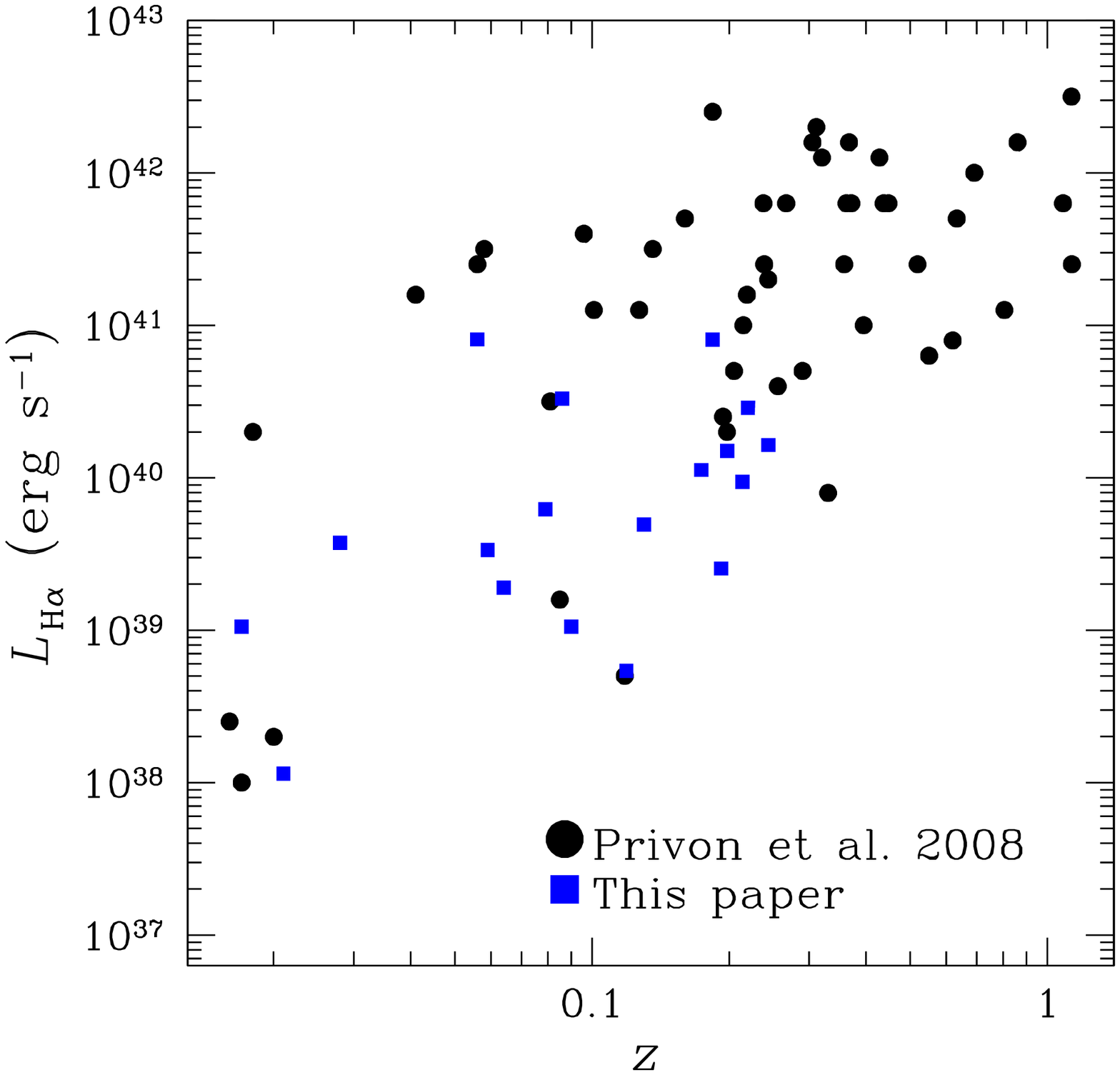}{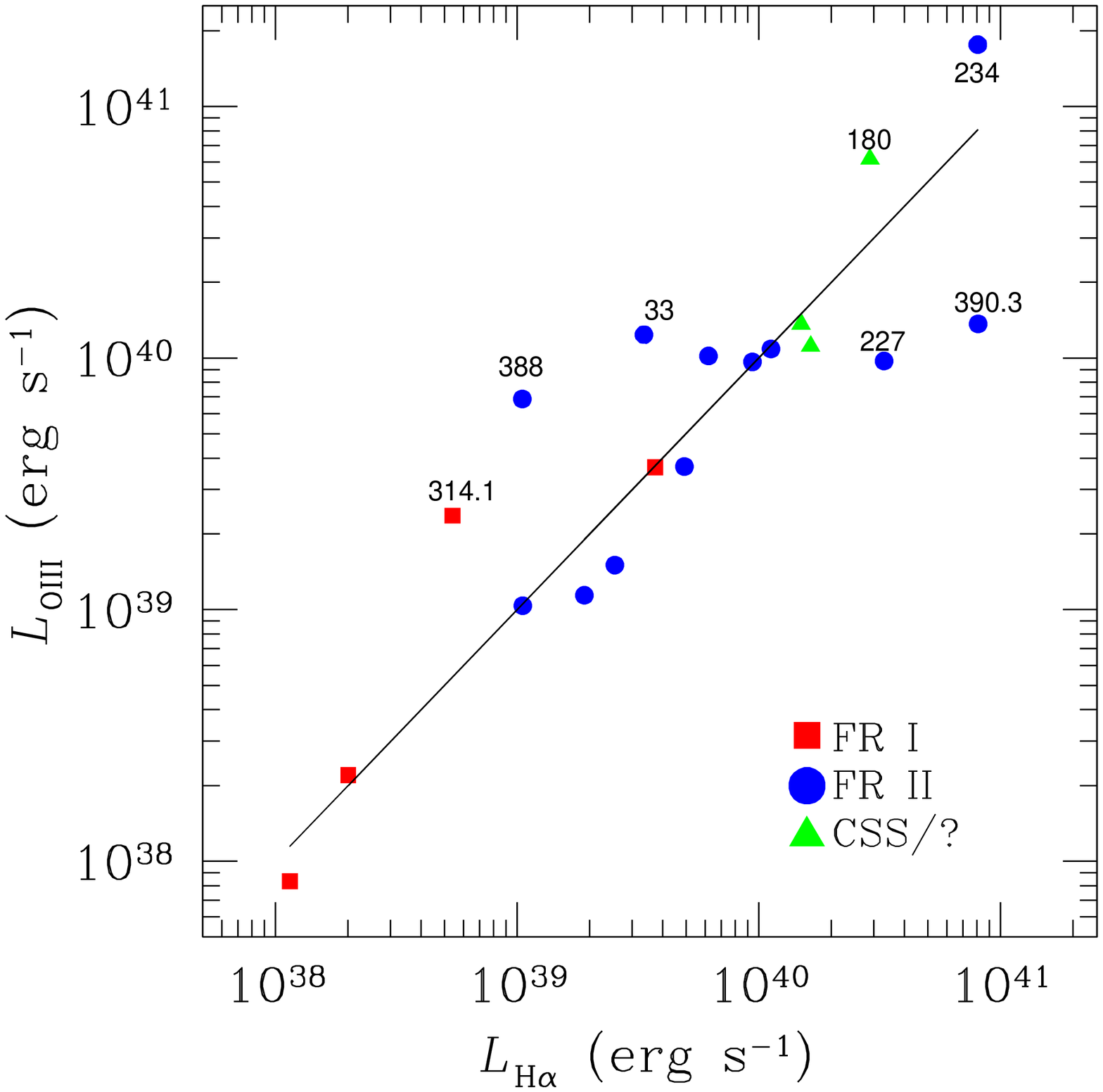}
\caption{ {\it  (Top left)} Measured \ha luminosity  (in erg s$^{-1}$)
  vs.   redshift for  the  19  galaxies in  our  sample.  Red  squares
  represent  the four  FR~Is in  our  sample, while  blue circles  are
  FR~IIs and green triangles  represent those objects exhibiting a CSS
  or  unclassified radio  morphology  (i.e., 3C~180,  see \S4.1.8  for
  details).  See \S3 for a details concerning our photometry strategy.
  {\it (Top right)} Measured  [O{\sc iii}] luminosities vs.  redshift,
  using the  same symbol  definitions as in  {\it (top  right)}. ({\it
    Bottom  left})  The  measured  \ha luminosities  from  our  sample
  (plotted in  blue squares) vs.  redshift, overplotted  with the {\it
    measured} and  {\it extrapolated} \ha luminosities  for the sample
  of 3CR  radio galaxies  from \citet{privon08}.  The  slightly larger
  scatter observed in the \citet{privon08} data reflects the fact that
  roughly  50\% of  their \ha  luminosities have  been  estimated from
  measured  [O{\sc  iii}]  luminosities  using  ratios  found  in  the
  literature.    In   general,   however,   our  galaxies   follow   a
  redshift-luminosity correlation  roughly consistent with  that found
  for a higher  redshift subset of the 3CR  from \citet{privon08}. See
  \S4.2 for  a discussion.  {\it  (Bottom right)} The  measured [O{\sc
      iii}]  luminosities vs.  the measured  \ha luminosities  for the
  objects in our sample. Those galaxies lying above the black diagonal
  line  (tracing  $L_{\mathrm{[OIII]}}  = L_{\mathrm{H}\alpha}$)  have
  [O{\sc iii}]  luminosities greater than their  \ha luminosities, and
  vice-versa. For  a discussion on the implications  of these emission
  line luminosity ratios, see \S4.2.  }
\label{fig:lumplots}
\end{figure*}

\subsection{Discussion: correlations}

Below we discuss significant correlations, consistent with the results
of previous related studies, that  are apparent in our data.  In \S4.3
we discuss  the physical interpretation of these  observed trends, and
compare our results to past literature.

\subsubsection{Emission line luminosities vs. redshift}

\begin{figure*}
\plotone{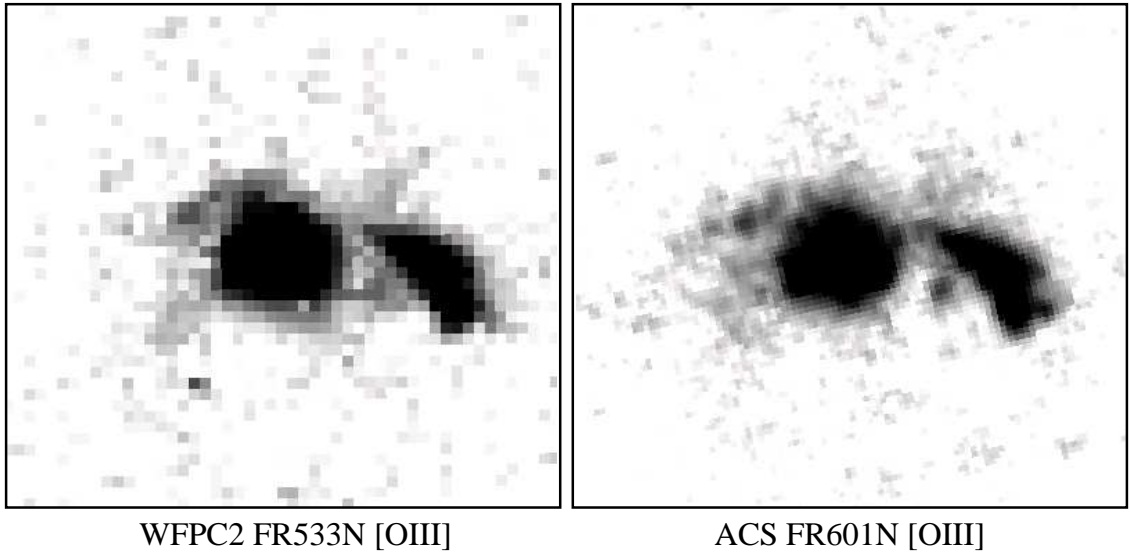}
\label{fig:comparison}
\caption{A representative  comparison of  data quality from  the WFPC2
  LRF  emission line  survey of  3CR  radio galaxies  as presented  in
  \citet{privon08}  in  comparison with  the  data  presented in  this
  paper.  ({\it a}) Continuum-subtracted WFPC2 FR533N $2 \times 300$ s
  exposure  of  the  [O{\sc  iii}]  emission  in  3C  234;  ({\it  b})
  continuum-subtracted  ACS WFC  FR601N exposure  of the  [O{\sc iii}]
  emission in 3C  234, from the dataset presented  in this paper.  See
  Table \ref{tab:tab2} and \S2 for details.  }
\label{fig:comparison}
\end{figure*}

In Fig.~\ref{fig:lumplots}{\it  a} and \ref{fig:lumplots}{\it  b} (top
left and top right, respectively)  we plot the measured \ha and [O{\sc
    iii}]      luminosities      ($L_{{\mathrm     H\alpha}}$      and
$L_{\mathrm{[OIII]}}$) vs.   redshift for the galaxies  in our sample.
FR~I  radio galaxies  are plotted  in  red squares,  while FR~IIs  are
plotted  in   blue  circles  and   the  three  galaxies   whose  radio
morphologies are  uncategorized are  plotted in green  triangles (this
convention holds  throughout this  paper).  As expected,  the emission
line luminosities follow  a positive redshift-luminosity trend, albeit
with  a greater degree  of scatter  than that  found in  the canonical
$K-z$  relation for  radio galaxies  (where  $K$ is  the NIR  $K$-band
magnitude of the elliptical radio galaxy host, see \citealt{willott03}
for  a detailed  study of  this correlation).   This higher  degree of
scatter is not necessarily surprising for a number of reasons, some of
which we summarize below:

\begin{enumerate}

\item Warm  optical line emitting  gas in radio galaxies  is naturally
  expected to inhabit a greater range in luminosity space than are the
  NIR  starlight  distributions of  their  elliptical  hosts.  As  the
  stellar component  of radio galaxies  is believed to have  formed at
  high redshift and evolved passively in the time since (barring other
  effects),  a tight  clustering of  $K$-band magnitudes  is expected.
  The  amount,  ionization state,  and  detectability  of  gas in  the
  nuclei  of radio  galaxies,  however,  is subject  to  a number  of
  factors that greatly influence overall emission line luminosities.

\item Our sample includes low-power FR~I radio galaxies, characterized
  by  narrow-line   emission,  and   FR~II  radio  galaxies   and  QSO
  characterized  by broad  optical  line emission.   Broad line  radio
  galaxies (BLRGs)  generally have an  optical excess, resulting  in a
  larger  contribution to  emission line  flux. Not  surprisingly, the
  BLRGs  in  our  sample  (3C~33,  197.1, 219,  285,  234,  \&  390.3)
  generally inhabit  the upper regions  of Fig.~\ref{fig:lumplots}{\it
    a}  and  \ref{fig:lumplots}{\it b},  contributing  to the  overall
  scatter.

\item These  galaxies range in redshift from $0.0075 <  z < 0.224$
  and were  observed with a  variety of LRF filter  configurations (see
  Table \ref{tab:tab2}).  This results in varying  amounts of emission
  line  flux  loss  beyond  the  edge  of the  passband,  as  well  as
  contamination from [N{\sc ii}]. 
\end{enumerate}

In Fig.~\ref{fig:lumplots}{\it c} we plot our \ha luminosities in blue
squares vs.   redshift, overplotted with  the {\it measured}  and {\it
  extrapolated} \ha luminosities for  the sample of 3CR radio galaxies
from \citet{privon08}.   The slightly  larger scatter observed  in the
Privon  data  reflects  the  fact  that  roughly  50\%  of  their  \ha
luminosities   have  been   estimated  from   measured   [O{\sc  iii}]
luminosities  using  ratios  found  in the  literature.   In  general,
however,  our  galaxies  follow  a redshift-luminosity  trend  roughly
consistent with that found  for the (generally) higher redshift subset
of the 3CR studied  in \citet{privon08}. See Fig. \ref{fig:comparison}
for a representative and qualitative  comparison of WFPC2 and ACS data
quality from the Privon observations and this paper, respectively.

\subsubsection{Emission line ratios}

In Fig.~\ref{fig:lumplots}{\it d} (bottom  right) we plot the measured
[O{\sc iii}]  luminosities vs.  the measured \ha  luminosities for the
objects in our sample.  Those  galaxies lying above the black diagonal
line  (tracing $L_{\mathrm{[OIII]}}  =  L_{\mathrm{H}\alpha}$) possess
[O{\sc  iii}] luminosities  greater than  their \ha  luminosities (and
vice-versa).   Note  that,  in  general, $  L_{\mathrm{H}\alpha}$  and
$L_{\mathrm{[OIII]}}$ are roughly similar across the sample, with some
notable exceptions whose names we have indicated in the figure.

In future  papers it  will be necessary  to investigate  emission line
ratio maps, which may carry important implications with regards to the
orientation-based   unification  scheme   of  radio-loud   AGN  (e.g.,
\citealt{padovani92,chiaberge04},  and  references  therein).   It  is
expected that low-excitation lines (like \ha) are emitted further from
the AGN  than are  higher excitation lines  like [O{\sc iii}],  as the
fraction of  highly excited gas will  depend on the  proximity of that
gas   to  the   ionizing  source,   in  this   case  the   AGN  (e.g.,
\citealt{jackson97}, and references therein).  This spatial dependence
of  $L_{\mathrm{[OIII]}}  /  L_{\mathrm{H}\alpha}$  also  leads  to  a
dependence upon the sources of  obscuration present in the galaxy.  In
the unification  model for radio  galaxies, wherein FR~Is  are unified
with BL-Lacs and FR~IIs are  unified as the parent population of steep
spectrum QSO,  we expect  low-excitation gas to  lie further  from the
obscuring torus,  while the  fraction of high-excitation  gas observed
will follow a tighter dependence  upon the angle at which the observer
views the obscuring torus.  We may therefore expect that intrinsically
similar radio-loud AGN viewed at a variety of angles will have unequal
[O{\sc   iii}]  luminosities  but   roughly  equal   \ha  luminosities
\citep{baum89b, jackson90}.

 It   will   also  be   important   to   investigate  whether   [O{\sc
     iii}]/H$\alpha$   ratios  of  order   unity  are   indicative  of
 Seyfert-like  nebular  activity.   We  note  that,  in  general,  the
 majority  of  objects  in  our  sample  with  $L_{\mathrm{[OIII]}}  /
 L_{\mathrm{H}\alpha}  \ne 1$ in  (i.e.  those  galaxies significantly
 offset from the diagonal  line in Fig.~\ref{fig:lumplots}{\it d}) are
 BLRGs, with the exception of one  FR~I.  Though we are unable to make
 statistically significant inferences with respect to these issues (as
 our sample consists of  only 19 galaxies), further investigation into
 these line ratios and associated  line ratio maps may yield important
 results.

\subsubsection{Emission line luminosities vs. total radio power}

\begin{figure*}
\plottwo{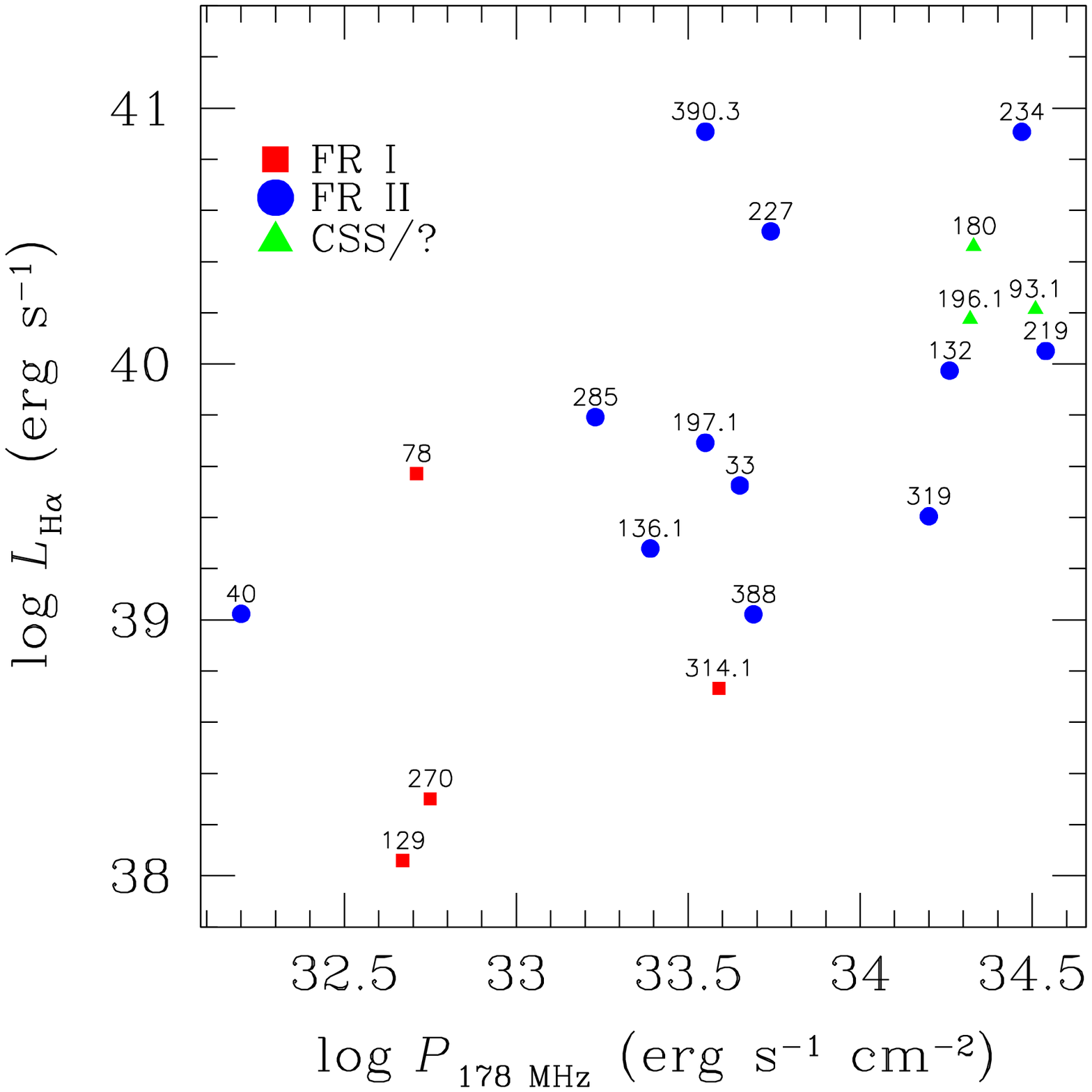}{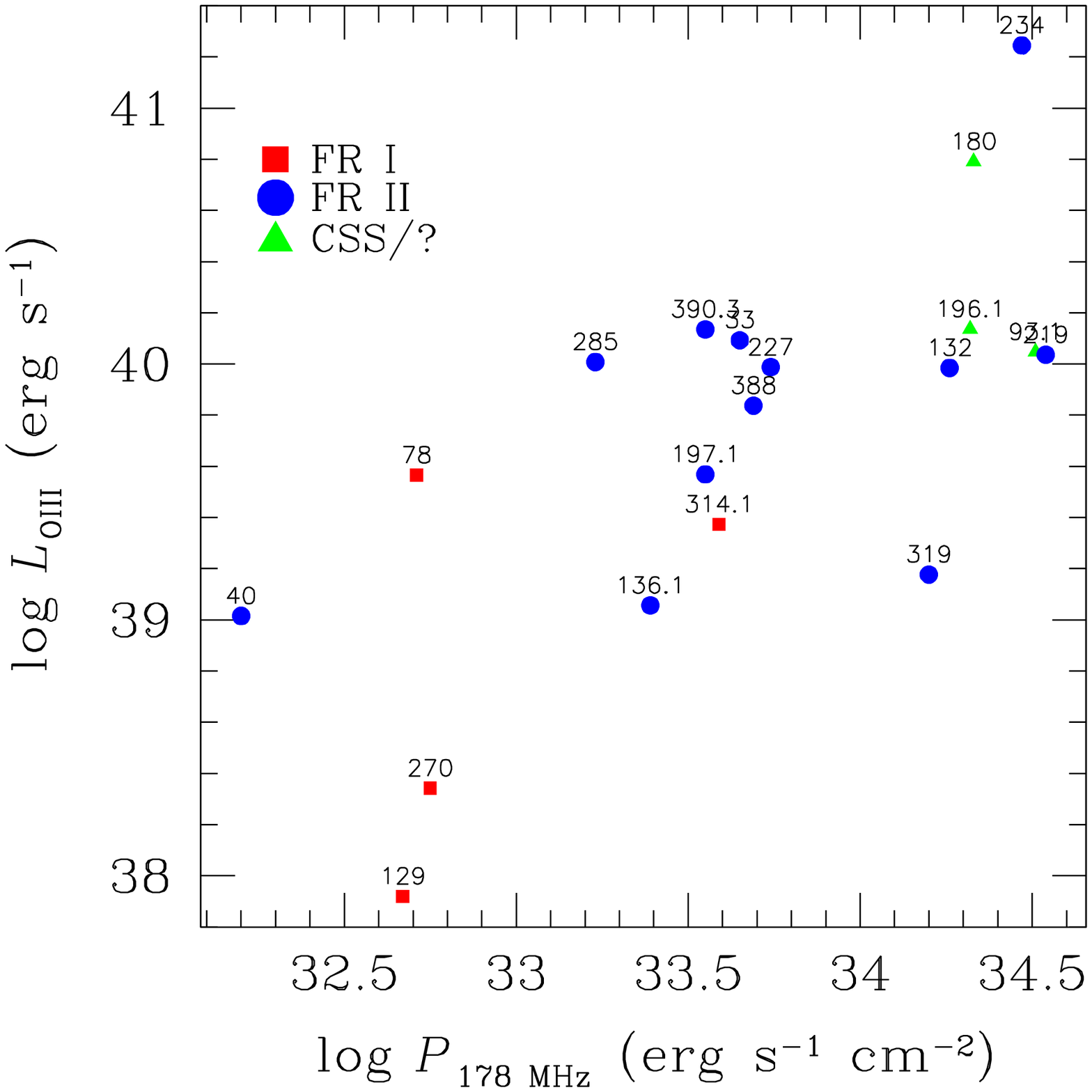}
\caption{Emission line luminosity vs. total radio power at 178 MHz for
  ({\it a})  \ha and ({\it b}) [O{\sc  iii}]. Radio luminosities
  are from \citet{spinrad85}. Red  squares represent the four FR~Is in
  our  sample,  while blue  circles  are  FR~IIs  and green  triangles
  represent  those  objects exhibiting  a  CSS  or unclassified  radio
  morphology.   We  note  the  clear  upward trend  in  both  figures,
  consistent  with   the  results   of  the  comprehensive   study  by
  \citet{willott99} which  found a tight  correlation between emission
  line  and  radio  luminosity   in  flux  limited  samples  of  radio
  galaxies. See Table \ref{tab:tab3} for a summary of the radio properties 
in our sample. }
\label{fig:radio}
\end{figure*}

In Figs.~\ref{fig:radio}({\it a}) and \ref{fig:radio}({\it b}) we plot
emission  line  luminosity vs.   total  radio  power  at 178  MHz  for
\ha and [O{\sc iii}], respectively. We note a clear upward trend
in  both figures,  consistent with  the results  of  the comprehensive
study  by  \citet{willott99}, which  also  found  a tight  correlation
between emission line and radio  luminosity in flux limited samples of
radio galaxies.

We  discuss interpretations  of this  result  in \S4.3, but  it is  first
important to frame  the results of the above  emission line luminosity
comparisons in the context of ionized gas distribution morphologies.

\subsubsection{Morphology of the extended narrow line regions}

The low-  and high- excitation  narrow-line regions we observe  in our
sample,   overall,   exhibit   extended   and   complex   morphologies
characterized  by (e.g.)   clumpy regions  of emission  (e.g.  3C~33),
whispy  tendrils  (3C~136.1,  3C~319),  extended filaments  and  lanes
(3C~196.1,  3C~234), and  bi-conical  series of  shells (3C~180).   In
general  the morphologies  observed in  [O{\sc iii}]  mirror  those in
\ha  with some important  exceptions (e.g.   3C~136.1, 3C~196.1,
3C~227,  3C~285).  We  also  note the  presence  of bright  unresolved
nuclear emission  in the cores of  every object with  the exception of
3C~129 in  [O{\sc iii}] and 3C~285  in \ha and  [O{\sc iii}] (no
clearly ``point  source-like'' nucleus is  apparent in our  images for
that object).

\begin{figure*}
\plottwo{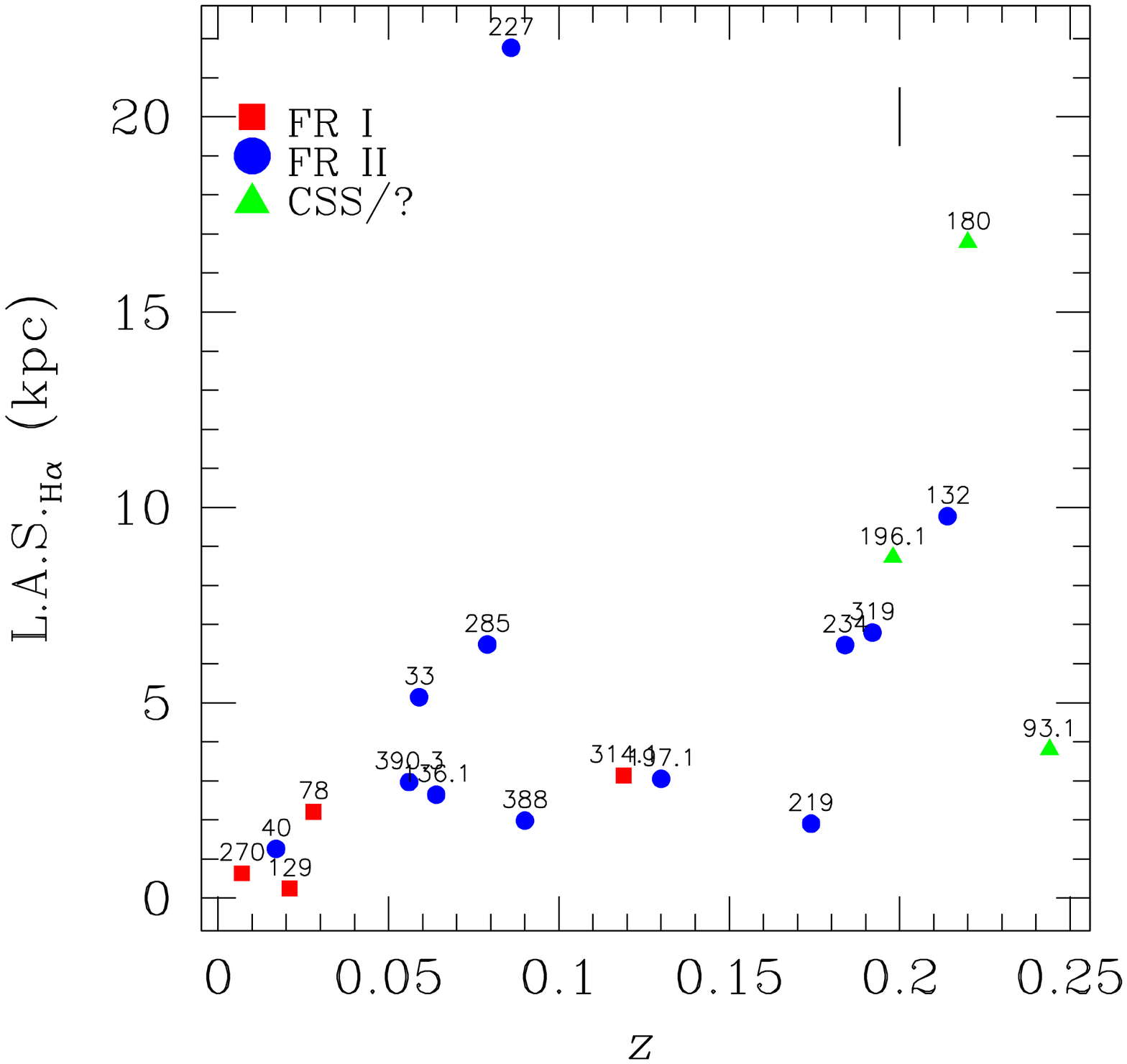}{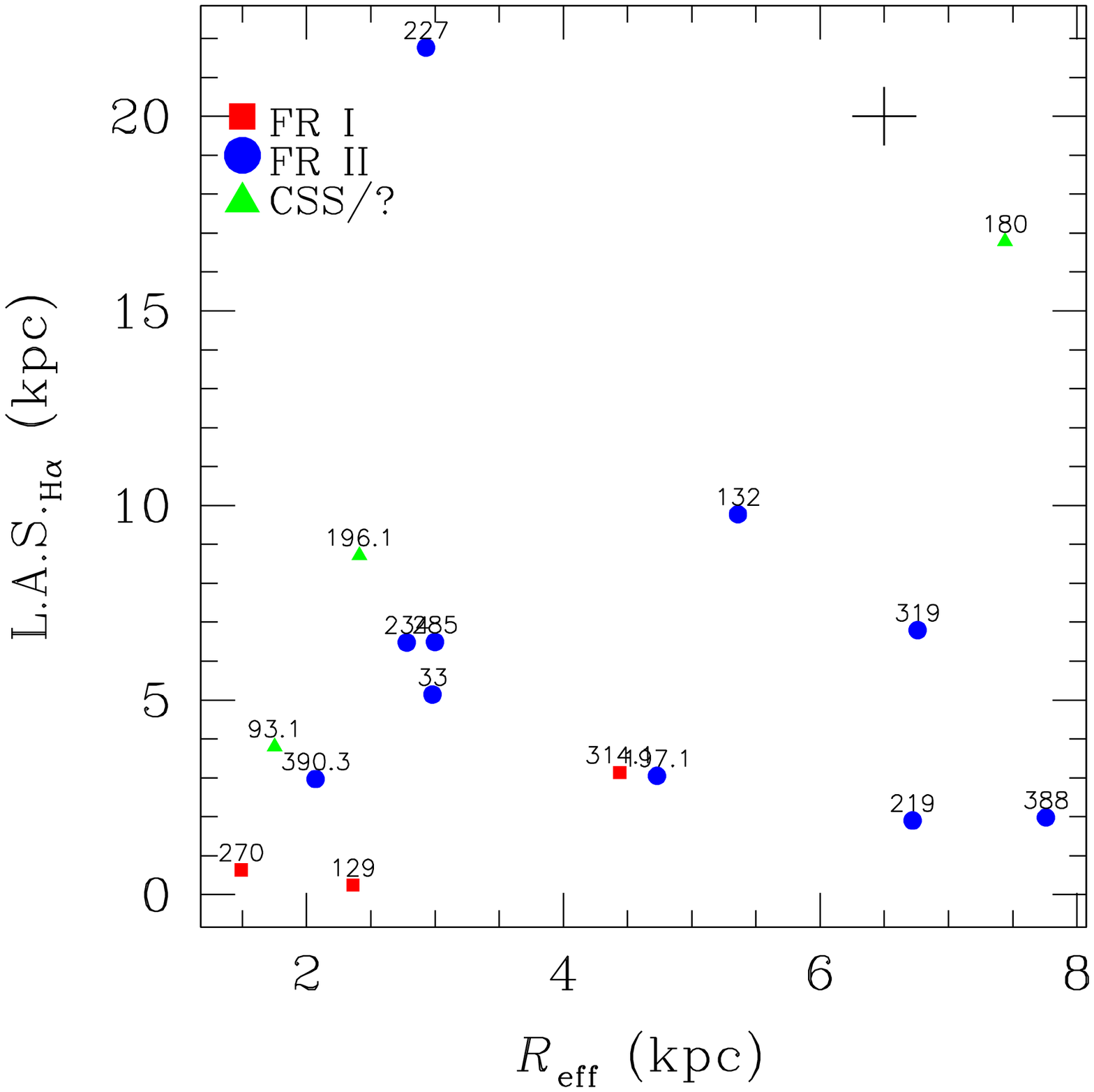}
\caption{Comparison of the projected linear sizes of the extended high
  surface brightness  \ha emission line regions  with both ({\it
    a}) redshift  and ({\it b}) effective  radii $R_{\mathrm{eff}}$ of
  the host galaxies.  The size estimates represent the largest angular
  size (L.A.S)  of the narrow  line regions detected in  our \ha
  imaging. We  do not plot the  L.A.S. of the [O{\sc  iii}] regions as
  their sizes  generally mirror those  of the \ha  regions, with
  some  important exceptions (see  \S4.2).  Characteristic  error bars
  appear in the upper right  corner of each plot. The effective radii,
  in  kpc, are  from \citet{floyd08}  and \citet{donzelli07}  and were
  derived from surface brightness  profile fitting to {\it HST}/NICMOS
  {\it  H}-band images  (and  some NIR  imaging  from {\it  Telescopio
    Nazionale  Galileo} in the  case of  \citealt{donzelli07}).  While
  the methods employed by these  two studies differ, their results are
  generally  consistent, and  any uncertainty  introduced  in plotting
  results from  both papers is  reflected in the  characteristic error
  bar we  have indicated. We  have omitted 3C~40, 3C~78,  and 3C~136.1
  from  ({\it b})  as  no estimate  for  their $R_{\mathrm{eff}}$  was
  available.  Note  the  general  upward  trend  of  the  L.A.S.  with
  redshift,    as   well    as    a   weaker    upward   trend    with
  $R_{\mathrm{eff}}$.  The  former   result  is  consistent  with  the
  emission line  study of \citet{privon08}, and is  expected given the
  tight redshift-luminosity correlation  resulting in the selection of
  higher power sources at  higher redshifts.  The higher-power sources
  are likely able to ionize  gas at greater distances from the central
  engine, contributing to the upward trend we observe. }
\label{fig:las}
\end{figure*}

In Figs.~\ref{fig:las}({\it a})  and \ref{fig:las}({\it b}) we compare
the  projected linear  sizes of  the line-emitting  regions  with both
redshift and  elliptical host galaxy  effective radius (respectively).
We have estimated the gas distribution sizes, in kpc, from the largest
angular  extent  (L.A.S.) subtended  by  the  high surface  brightness
emission (as projected  on the sky).  While this  is a rough estimate,
it provides important  insights as evidenced by the  weak upward trend
of  L.A.S.  with redshift  (Fig.~\ref{fig:las}{\it a}),  implying that
larger distributions of optical line  emitting gas are found at higher
redshifts.

Values    for    the    effective    radii    $R_{\mathrm{eff}}$    in
Fig.~\ref{fig:las}({\it    b})    are    from   \citet{floyd08}    and
\citet{donzelli07}, and  were derived from  surface brightness profile
fitting to {\it HST}/NICMOS {\it  H}-band images (and some NIR imaging
from TNG,  in the case  of \citealt{donzelli07}).  While  the methods
employed  by these  two studies  differ, their  results  are generally
consistent, and  any uncertainty  introduced in plotting  results from
both  papers is  reflected in  the characteristic  error bars  we have
indicated  in the  figure.   In light  of  these added  uncertainties,
however,  we are  unable to  make any  significant  inferences, though
L.A.S.   and  $R_{\mathrm{eff}}$   do  appear  weakly  correlated  for
galaxies with $R_{\mathrm{eff}} < 4$ kpc, while no such correlation is
evident for those objects with $R_{\mathrm{eff}} > 4$.

\subsection{Physical interpretation of the trends observed}

We have discussed three trends apparent in our data:

\begin{enumerate}

\item Emission  line luminosity (both  for \ha and  [O{\sc iii}])
  appears positively correlated with the redshift of the host galaxy.

\item  The \ha and  [O{\sc iii}]  luminosities are  also tightly
  correlated  with  total radio  power  at  178 MHz.

\item Projected sizes  of the ionized gas distributions  in our sample
  are positively  correlated with host galaxy redshifts.   That is, we
  observe larger line-emitting regions at greater distances.

\end{enumerate}

The   above   results   are    consistent   with   those   of   (e.g.)
\citet{willott99,  willott03, privon08}  (and references  therein), as
well  as  many  other past  studies  of  emission  line gas  in  radio
galaxies.  The  tight redshift-luminosity correlation  in flux-limited
samples results in the  selection of intrinsically brighter objects at
higher redshifts. In  radio flux density selected samples  such as the
3CR,  this  is  manifest  in  more powerful  radio  galaxies  (FR~IIs)
inhabiting  a  higher  redshift  range  than  do  their  intrinsically
lower-power  counterparts  (FR~Is).   Moreover, the  alignment  effect
discussed in  \S1 suggests that propagation  of the radio  jet plays a
role   in   the   excitation   of   the   ISM   into   line   emission
\citep{baum89a,mccarthy93,best00}.  It  is therefore unsurprising that
more powerful  radio galaxies  appear associated with  higher emission
line  luminosities,  as they  are  able  to  collisionally excite  and
photoionize greater amounts of gas than are lower-power (and therefore
lower-redshift)                                                     AGN
\citep{baum89a,baum89b,rawlings91,inskip02a,inskip02b}.   As accretion
disk radiative  luminosity appears related  to the jet  kinetic energy
flux \citep{baum89b,rawlings91}, this can be used to place constraints
on the  physics of the AGN.   The greater mechanical  energy input and
number  of ionizing photons  provided by  more powerful  radio sources
also explains the dependence of emission line region size on redshift:
we expect the  more powerful radio galaxies at  higher redshifts to be
capable of exciting gas at  greater distances from the central engine,
resulting in the observed trend.


\begin{deluxetable}{llccccc}
\tablecaption{Radio Properties}
  \tablewidth{0pc}
  \tablehead{
    \colhead{} & 
    \colhead{} &
    \colhead{$S_{178~\mathrm{MHz}}$   } &
    \colhead{log $P_{178~\mathrm{MHz}}$  }\\
    \colhead{Source} &
    \colhead{Type} &
    \colhead{(Jy)} &
    \colhead{(erg s$^{-1}$ Hz$^{-1}$) } & \\
    \colhead{(1)} & \colhead{(2)} & \colhead{(3)} &
    \colhead{(4)}  }
  \startdata                                                                                     
3C~33.0  & FR~II &49.0   &33.65   \\
3C~40.0  & FR~II & 24.0  & 32.20     \\
3C~78.0  & FR~I &15.0 & 32.71        \\
3C~93.1  & CSS &10.0  & 34.51        \\
3C~129.0 & FR~I &21.0   &32.67      \\
3C~132.0 & FR~II  & 12.5  &34.26      \\
3C~136.1 & FR~II & 13.0   &  33.39     \\
3C~180.0 & \nodata &14.0   &34.33      \\
3C~196.1 & CSS  & 14.5   &34.32       \\
3C~197.1 & FR~II &9.5    &33.55      \\
3C~219.0 & FR~II & 44.0   &34.54     \\
3C~227.0 & FR~II & 28.0  &33.74     \\ 
3C~234.0 & FR~II  &29.0   &34.47      \\
3C~270.0 & FR~I & 44.0    &32.75      \\
3C~285.0 & FR~II  & 10.5  &33.23     \\
3C~314.1 & FR~I &  9.0   &33.59      \\
3C~319.0 & FR~II  & 17.0  &34.20      \\
3C~388.0 & FR~II & 22.0   &33.69      \\
3C~390.3 & FR~II & 44.0   &33.55      \cr
  \enddata
  \tablecomments{
    (1) Source name;
    (2) Fanaroff-Riley classification; 
    (3) 178 MHz flux density in Jy;
    (4) Log$_{10}$ of radio power at 178 MHz, in erg s$^{-1}$. 
  }
  \tablerefs{\citealt{fanaroff74, spinrad85}}
\label{tab:tab3}
\end{deluxetable}

It  should also  be noted  that dense  clouds of  gas are  observed on
increasingly larger scales at higher redshifts, ultimately forming the
vast Ly-$\alpha$ halos that  surround the most powerful radio galaxies
\citep{vanojik97,reuland03}.   While  this  dependence  of  gas  cloud
density and extent on redshift is important for objects with $z > 0.6$
(the point near which the alignment effect precipitously tightens), it
is not likely to contribute in any meaningful way to our comparatively
low redshift sample at $z < 0.3$.

\section{Summary and Concluding Remarks}

In this paper  we have presented {\it HST}/ACS  narrow-band imaging of
\ha and [O{\sc iii}]$\lambda$5007 line-emission in the central regions
of  19 low-redshift  radio galaxies  from  the 3CR  catalog.  We  have
chosen the 3CR as the basis for our sample as it is radio flux density
selected  at a  frequency dominated  by unbeamed  radio lobes,  and is
therefore free  from orientation  bias with respect  to all  {\it HST}
wavelengths.  Moreover,  the 3CR has been extensively  covered by past
ground- and space-based imaging  and spectroscopy programs, yielding a
robust, cross-spectrum database  containing galaxies exhibiting a wide
variety  of intrinsic  characteristics.  We  limit our  sample  to the
low-redshift  subset  ($z <  0.3$)  of  the  3CR to  maximize  spatial
resolution, as well  as to ensure that the  emission lines fall within
the  redshift range of  high sensitivity.   The ACS  observations were
carried out during {\it HST} Cycle  15, and in total, 19 galaxies were
observed  prior  to  failure   of  the  instrument  in  January  2007.
Reduction of  these data was performed independently  of the automated
OTFR  pipeline  so  as  to   improve  on  flat-fielding  and  to  more
proactively  clean the  images  of  cosmic rays  and  hot pixels.   In
addition,  contribution   from  continuum  was   subtracted  to  yield
effectively pure emission line images.

We have  compared the resulting  emission line luminosities  with both
redshift and total radio power, finding weak positive correlations for
each.  We have  also noted an upward trend  between the projected size
of the  emission line gas  distributions and redshift.   These results
are consistent with  those from past studies, and  we have interpreted
the apparent  trends to  be a result  of higher power  radio galaxies,
preferentially  found  at  greater  distances,  which  can  shock  and
photoionize greater amounts of the  ISM further away from the nucleus,
thus resulting in brighter and more extended emission-line regions.

Regardless of  the interpretation,  recombination times for  warm gas
($T  \sim 10^4$  K) are  of order  $10^3$  years \citep{osterbrock89},
while  lifetimes of  radio sources  are  of order  $10^8$ years.   The
process  by which  the  gas  is ionized  should  therefore be  ongoing
throughout the  lifetime of the  source, implying a  strong connection
between  AGN  activity and  observed  emission  line properties.   The
sensitive, high spatial resolution  images presented in this paper may
therefore enrich  future works  studying jets propagating  through the
ISM, and  their relationship with extended and  compact star formation
regions as well  as X-ray coronal halos. Future  emission line studies
of larger samples  of radio galaxies may also  deepen understanding of
radio-loud  unification schemes,  as  warm optical  line emitting  gas
traces fundamental  energy transport processes coupled  to feedback at
the interface of the AGN and its surrounding medium.

\acknowledgments  We  are  indebted  to the  anonymous  referee  whose
comments led to the improvement  of this work.  We thank George Privon
for  helpful discussions  and access  to  data.  G.~R.~T.~acknowledges
A.~N.~S.~and support from NASA grant HST-GO-10882.01-A, as well as the
New York Space  Grant Consortium.  This research made  use of the NASA
Astrophysics  Data  System bibliographic  services  and the  NASA/IPAC
Extragalactic  Database, operated  by the  Jet  Propulsion Laboratory,
California Institute of Technology, under contract with NASA.

\bibliography{ms_rev1.bbl}

\end{document}